\documentclass[runningheads]{llncs}
\usepackage[T1]{fontenc}
\usepackage{graphicx,verbatim}
\usepackage{booktabs}
\usepackage{multirow}  
\usepackage{stmaryrd} 
\usepackage{amssymb}
\usepackage{amsmath}
\usepackage{siunitx}
\usepackage[table]{xcolor}

\usepackage[colorlinks=true,
            linkcolor=blue,     
            citecolor=blue,     
            urlcolor=blue
           ]{hyperref}

\begin{document}
%
\title{TRUE-Colon: Exposing a Consistent Transfer Asymmetry in Real-Time Polyp Detection}
%
\titlerunning{Exposing a Consistent Transfer Asymmetry in Polyp Detection}
%
\author{Sebastian Doerrich\thanks{These authors contributed equally to this work.} \and
Andreas Franz Schwab\protect\footnotemark[1] \and
Francesco Di Salvo \and
Shyam Nandan Rai \and
Hanh Huyen My Nguyen \and
Christian Ledig}
\authorrunning{S. Doerrich et al.}
%
\institute{xAILab Bamberg, University of Bamberg, Bamberg, Germany
\email{sebastian.doerrich@uni-bamberg.de}}
\maketitle              
\begin{abstract}
Computer-aided detection (CADe) systems for colonoscopy promise to reduce clinical miss rates, yet reliable real-world deployment remains elusive. This translational gap stems in part from a structural flaw in model development: the reliance on curated datasets that under-represent the long negative stretches and procedure-related artifacts characteristic of routine examinations. Training and evaluating architectures strictly on these lesion-centric benchmarks creates an illusion of success, since such benchmarks cannot capture clinically crucial metrics. To expose this gap, we establish TRUE-Colon, a standardized benchmarking protocol that measures key deployment characteristics alongside localization accuracy, and evaluate four real-time architectures (Faster R-CNN, YOLOv8, YOLOv11, RT-DETR) across curated benchmarks (SUN, PICCOLO) and 60 unedited, full-length procedures (REAL-Colon). We observe a consistent transfer asymmetry: models trained strictly on curated clips suffer a severe performance collapse when evaluated on full procedures, whereas procedure-trained models substantially improve rejection of non-polyp content on REAL-Colon, and largely retain their accuracy on curated benchmarks. Beyond transferability, we find that the Transformer detector attains the strongest sensitivity and the earliest, most persistent detections, while the convolutional detectors stay competitive at a higher throughput. Together, these results indicate that both training and benchmarking for deployable CADe should shift from curated, lesion-centric clips toward full-procedure data and deployment-relevant operating points. Source code is available at \url{https://github.com/sdoerrich97/true-colon}.

\end{abstract}
\section{Introduction}
\begin{figure}[h]
    \centering
    \setlength{\tabcolsep}{2pt}
    \renewcommand{\arraystretch}{1.1}
    \begin{tabular}{c c c c c}
        \includegraphics[width=0.185\linewidth]{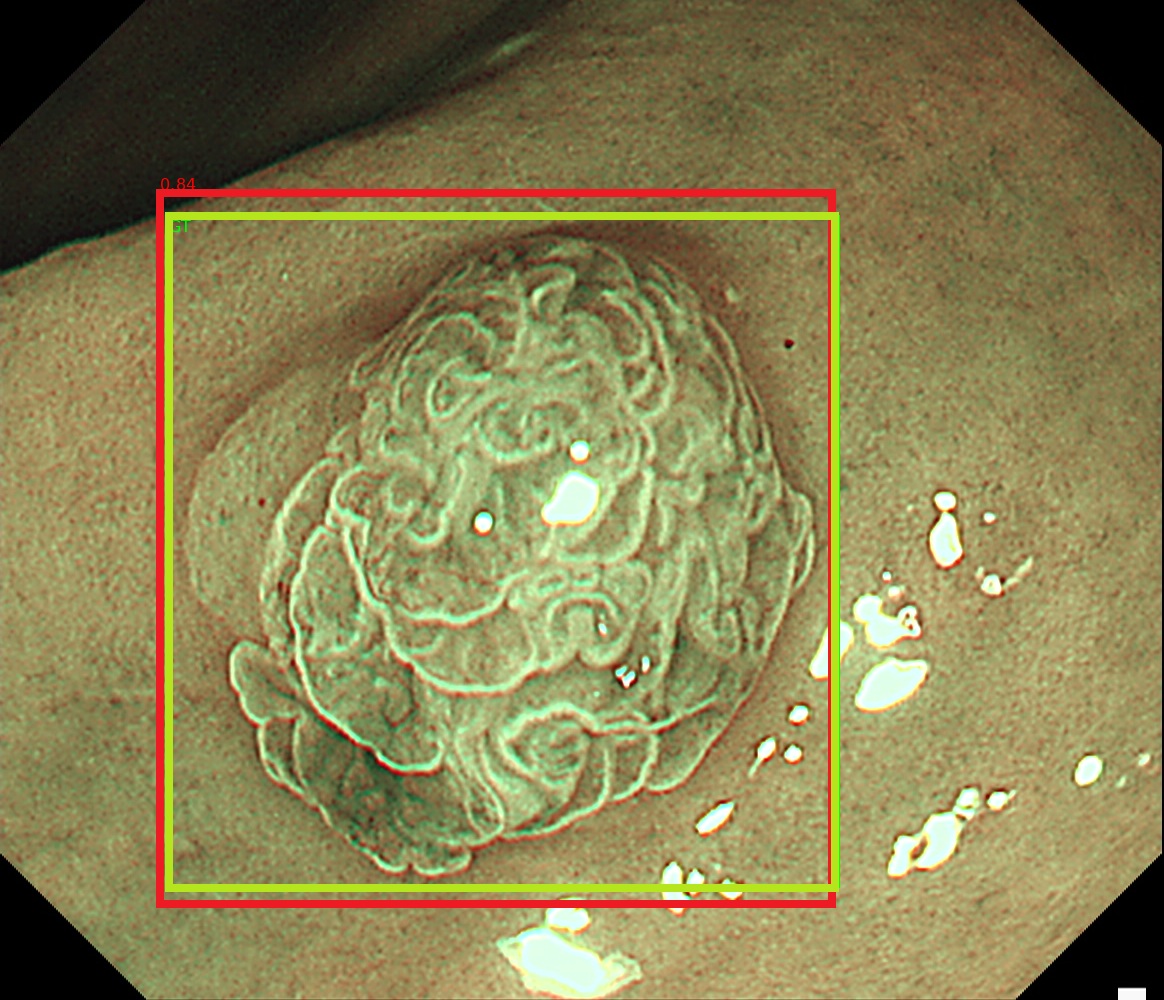} &
        \includegraphics[width=0.185\linewidth]{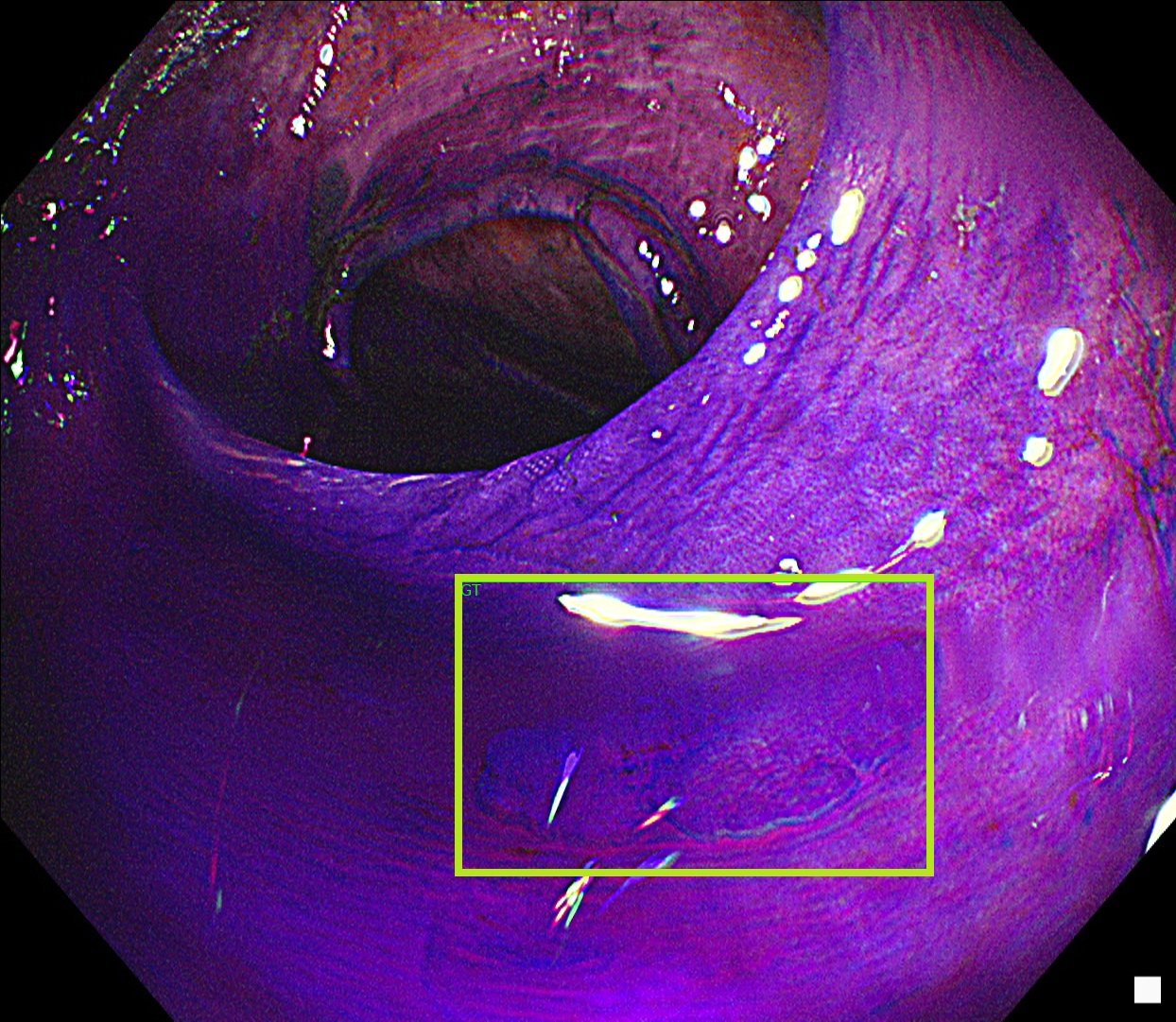} &
        \includegraphics[width=0.20\linewidth]{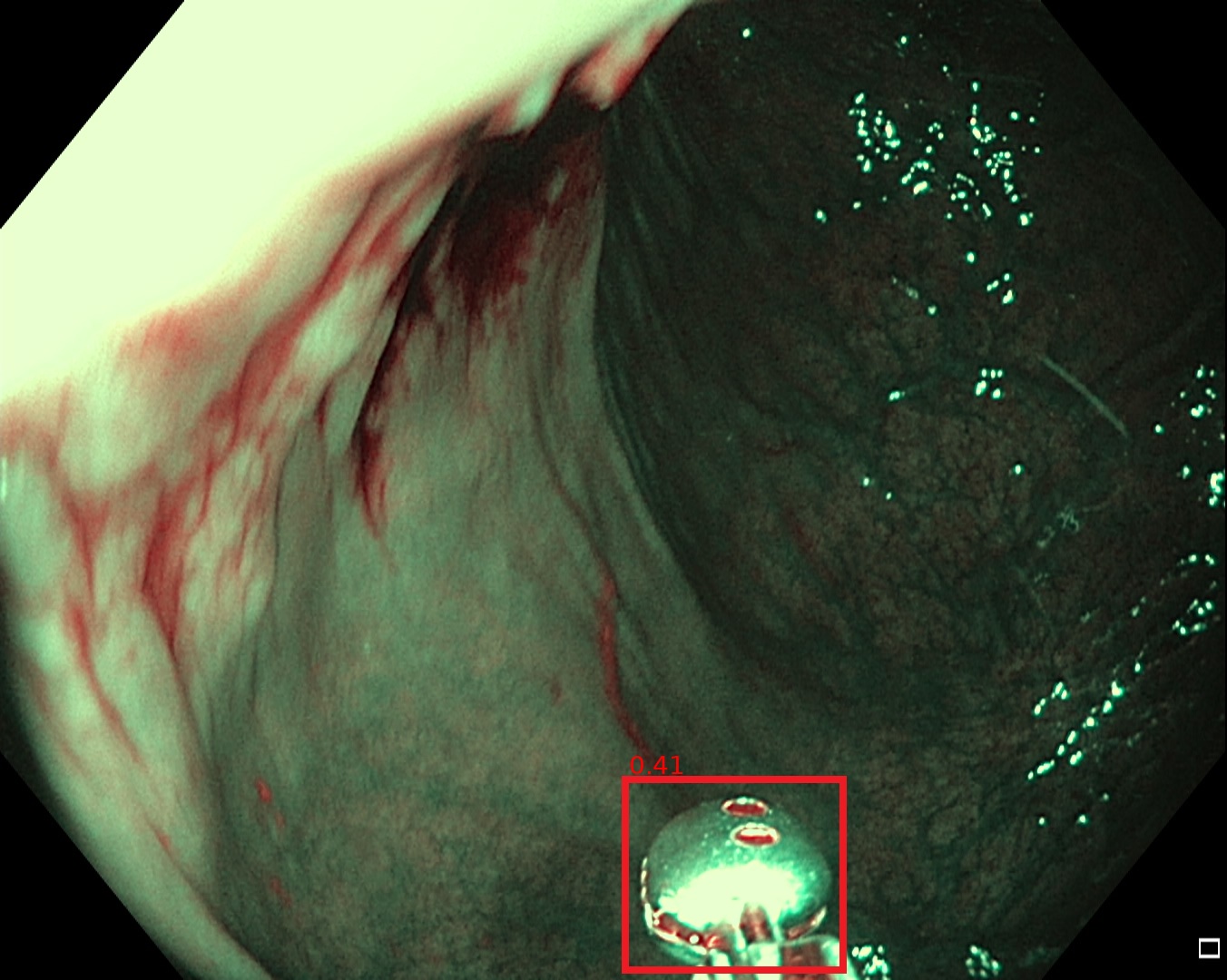} &
        \includegraphics[width=0.185\linewidth]{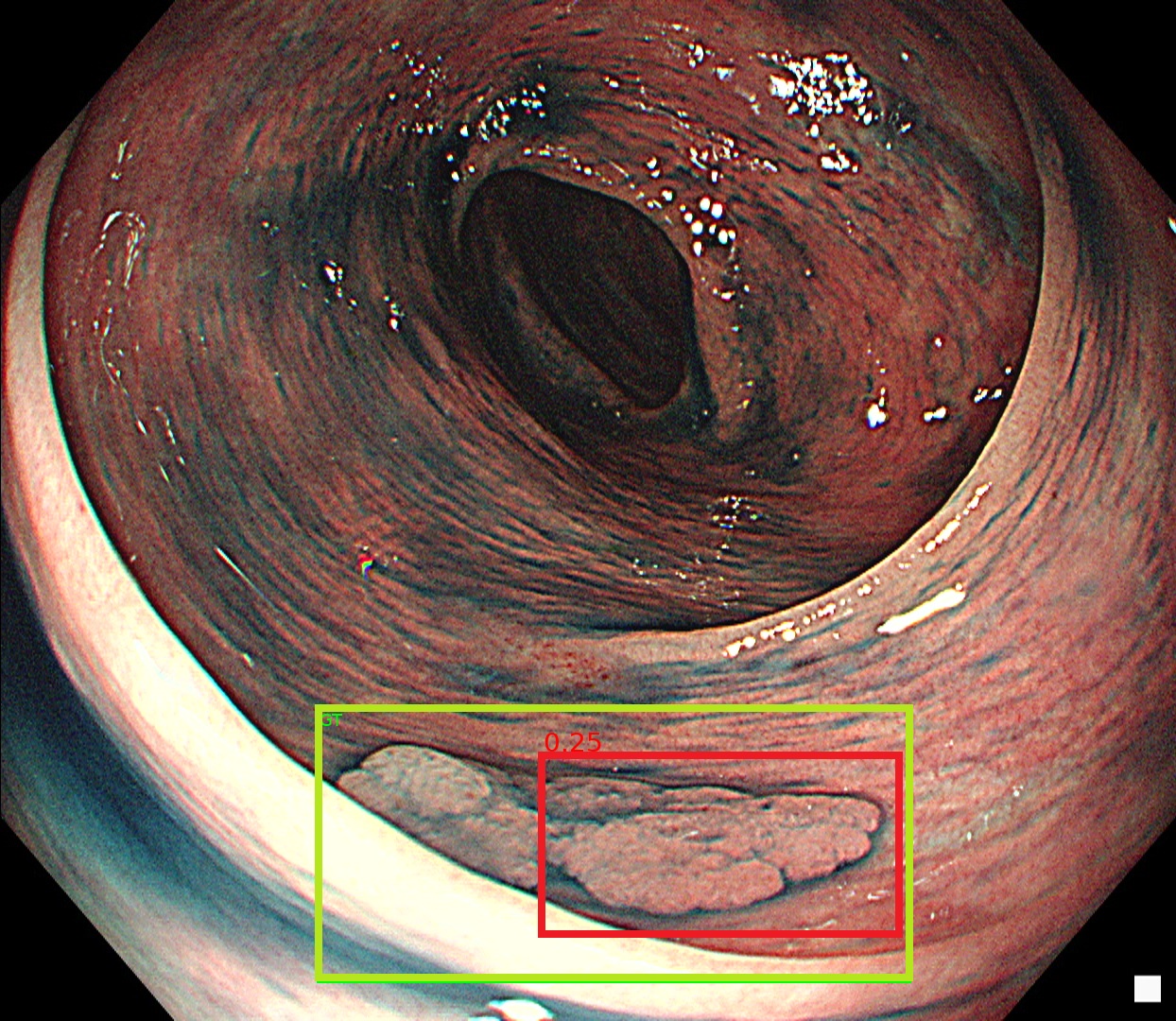} &
        \includegraphics[width=0.185\linewidth]{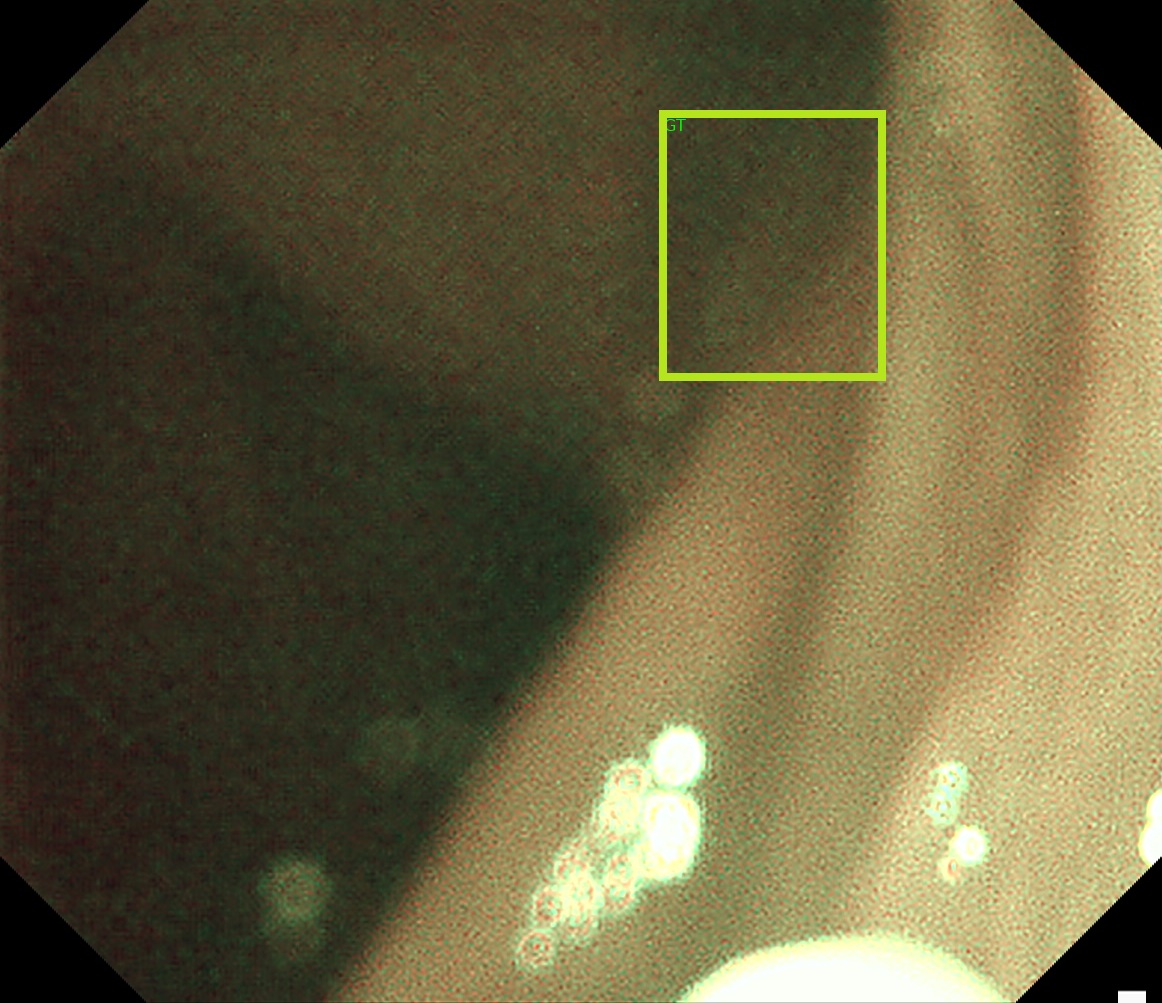}
    \end{tabular}
    \caption{Qualitative YOLOv11 results on representative colonoscopy frames. The examples illustrate varying levels of difficulty, including clear polyp visibility, specular highlights, motion blur, and challenging background artifacts. Green bounding boxes show the ground-truth, while red bounding boxes indicate predictions.}
    \label{fig:qualitative_overview}
\end{figure}

Colonoscopy is the clinical gold standard for colorectal cancer prevention, relying on the real-time detection and resection of precursor lesions~\cite{new_england}. In practice, however, visual inspection occurs in a complex, continuous endoscopic environment. Polyps appear transiently, vary strongly in morphology, and are frequently obscured by folds, debris, and changing illumination~\cite{bg_content_based_processing}. These challenges contribute to clinical miss rates of up to $27\%$, driving the critical need for computer-aided detection (CADe) systems to act as an automated safety layer~\cite{polyp_miss_rate}.
Modern CADe systems rely on deep neural object detectors to meet real-time, high-sensitivity requirements. Despite high accuracy on public benchmarks, clinical deployment remains restricted. This translational gap exists because standard evaluation protocols and reference datasets heavily curate their data around lesion-centric clips~\cite{trans_gap}. Such curated distributions under-represent the long stretches of negative frames (i.e., frames without an annotated polyp) and procedure-related artifacts, such as specular highlights, motion blur, and instruments, that characterize full clinical examinations. Evaluating architectures strictly on these datasets therefore conceals the false-alert burden they trigger, creating a dangerous illusion of success (\figurename~\ref{fig:qualitative_overview}).
Furthermore, the academic literature is methodologically inconsistent. Prior studies~\cite{lee_rt,kumc,endomind,related_yolo,chr_yolo,yoo_1,deformable_detr} use disjoint protocols with non-standardized data splits, IoU thresholds, and temporal handling, complicating comparisons of relevant behavior beyond localization. We address this by establishing TRUE-Colon, a standardized benchmarking protocol on REAL-Colon~\cite{realColon} (60 full, unedited examinations), which quantifies localization accuracy, false-alert burden, detection latency, and temporal reliability at a matched false-alert operating point. Across curated benchmarks (SUN, PICCOLO) and full procedures, we observe a consistent transfer asymmetry: models trained on curated clips degrade sharply on full procedures, whereas procedure-trained models learn robust suppression of non-polyp content and generalize well back to curated benchmarks.
The main contributions of this work are:
\begin{itemize}
    \item We establish TRUE-Colon, a standardized protocol for deployment-relevant evaluation. Alongside localization accuracy, it quantifies false-alert burden, detection latency, and temporal reliability, at a matched false-alert operating point that enables direct comparison across architectures.
    \item We expose a consistent transfer asymmetry across four detectors: curated-trained models collapse on full procedures, whereas procedure-trained models transfer well to curated benchmarks, showing that full-procedure data are the more informative training signal for deployment.
    \item We find that at a matched false-alert operating point, the Transformer detector leads on sensitivity, latency, and temporal persistence, while the convolutional detectors remain competitive at substantially higher throughput, an accuracy-versus-compute trade-off that fixed-threshold comparisons obscure.
\end{itemize}
\section{Methodology}
\label{sec:Methodology}
\subsection{Evaluation Perspective and Deployment Characteristics}
\label{sec:metrics}
Standard object detection metrics quantify localization quality on annotated instances, but they do not capture clinically critical behavior in continuous colonoscopy videos. Therefore, we define and report four complementary characteristics throughout the paper: (i) localization accuracy, (ii) false-alert burden, (iii) detection latency, and (iv) temporal reliability. For all sequence-based evaluations, we define a frame as \emph{positive} if it contains at least one annotated polyp.

\subsubsection{Localization Accuracy (Instance Localization).}
We evaluate how accurately the model localizes individual polyps by matching predictions to ground-truth bounding boxes using standard Intersection-over-Union (IoU) criteria. We report mean Average Precision $\mathrm{mAP}_{50}$ (IoU $\geq 0.50$) and $\mathrm{mAP}_{50{:}95}$ (averaged over IoU thresholds from $0.50$ to $0.95$). While these metrics isolate spatial accuracy, they do not quantify operational false alerts on negative video segments.

\subsubsection{False-Alert Burden (Frame-Level Alert Behavior).}
To translate detections into clinical alert behavior, we collapse detector outputs into a binary decision (alert vs.\ no alert) at a specified confidence threshold $\tau$. A detection counts as \emph{valid} when its confidence exceeds $\tau$ and, on a positive frame, overlaps an annotated polyp box (IoU $>0$). A frame is a \emph{true positive} if it is positive and the model produces at least one valid detection; a \emph{false negative} if it is positive but no valid detection is produced. A negative frame triggers a \emph{false positive} if the model raises at least one alert. This yields the True Positive Rate (TPR) and False Positive Rate (FPR), which we use as a measure of false-alert burden.

\subsubsection{Detection Latency (Time-to-First Detection).}
Clinical usefulness also depends on how early a detector fires once a lesion becomes visible. For each lesion, we measure latency as the number of frames from the lesion's annotated onset to the first valid detection. Because the detectors are memoryless and operate per frame, this quantity reflects sensitivity on early, partially visible frames rather than a temporal reaction speed, and is coupled to the chosen operating point. We therefore report it at the matched $\tau^\star$ alongside frame-level sensitivity.

\subsubsection{Temporal Reliability (Lesion-Level Consistency).}
Finally, we quantify temporal reliability over a lesion's visible time window: a lesion is \emph{detected} if the model produces at least one valid detection during this window, and \emph{persistent} if valid detections occur in at least half of its visible frames, thereby distinguishing sporadic detections from consistent tracking. Lesion-level temporal metrics require continuous videos with consistent per-lesion identifiers (or an equivalent definition of a lesion's visible time window). These requirements are met by REAL-Colon, but are not available in SUN and PICCOLO. Consequently, we report temporal reliability and latency only on REAL-Colon.

\subsection{Cross-Distribution Transfer Analysis}
\label{subsec:cross_distribution_transfer}
To assess how dataset realism affects these characteristics, we evaluate cross-distribution transfer. Let $\mathcal{D}_{\text{proc}}$ denote a continuous procedure-level distribution (REAL-Colon) and $\mathcal{D}_{\text{cur}}$ denote curated benchmark distributions (SUN, PICCOLO). We evaluate four settings: in-domain curated ($\mathcal{D}_{\text{cur}} \rightarrow \mathcal{D}_{\text{cur}}$), curated-to-procedure transfer ($\mathcal{D}_{\text{cur}} \rightarrow \mathcal{D}_{\text{proc}}$), in-domain procedure ($\mathcal{D}_{\text{proc}} \rightarrow \mathcal{D}_{\text{proc}}$), and procedure-to-curated transfer ($\mathcal{D}_{\text{proc}} \rightarrow \mathcal{D}_{\text{cur}}$). This design tests whether full-procedure training improves robustness on continuous videos while maintaining benchmark performance.

\subsection{Training and Inference Setup}
\label{sec:train_infer_setup}
To separate data realism effects from architectural variance, we benchmark four real-time detectors: Faster R-CNN~\cite{faster_rcnn}, YOLOv8-M~\cite{yolo8}, YOLOv11-M~\cite{yolo11}, and RT-DETR~\cite{rtdetr_original}. We train all models using their default recipes for 100 epochs with early stopping (patience 10). Faster R-CNN utilizes \texttt{Detectron2}~\cite{detectron2} (effective batch size 96), while the remaining architectures use \texttt{Ultralytics}~\cite{yolo8,yolo11,rtdetr_ultralytics} (effective batch size 208). All networks process a fixed $640\times640$ input resolution, and we aggregate results across three random seeds to account for training variance.
During inference, we apply a low-confidence pre-filter ($\text{conf}=0.001$) to preserve raw proposal distributions for operating-point analysis. For REAL-Colon frame and lesion evaluations, we report results at a fixed threshold ($\tau=0.2$) to maintain comparability with prior work~\cite{endotest,endomind}. In addition, to compare models at an equalized false-alert burden, we also report results at a matched operating point $\tau^\star$ chosen \emph{post-hoc} such that the \emph{evaluated} frame-level false-positive rate on the REAL-Colon test set is approximately $\mathrm{FPR}\approx 4$--$5\%$ (following~\cite{URBAN20181069}). This normalization is used solely for controlled comparisons across models with different score calibration and should not be interpreted as an operating-point selection procedure for deployment.
Finally, to prevent artificial latency inflation from annotation discontinuities, we exclude lesions with an annotation gap exceeding 50 frames within their first 250 frames of visibility.
\section{Experiments and Results}
\subsection{Datasets and Partitioning}
We evaluate performance across three dataset distributions (\tablename~\ref{tab:datasets}). REAL-Colon~\cite{realColon} provides procedure-level realism via 60 full, unedited colonoscopy videos, with strict patient-level independence and a $10/2/3$ (train/val/test) video split per institution. SUN~\cite{sun} is a curated, lesion-centric video benchmark with $70/10/20$ positive clips and $7/2/4$ negative clips (train/val/test). PICCOLO~\cite{piccolo} is a highly curated still-image benchmark, for which we use the official author-provided splits (images and anonymized data from patients included in this study were provided by the PICCOLO database from the Basque Biobank \url{www.biobancovasco.bioef.eus}).
\begin{table}[h]
    \caption{Dataset distribution including positive/negative frame counts per split and proportion of negative frames. REAL-Colon maintains the highest proportion of negative frames ($>85\%$) in contrast to the curated benchmarks SUN and PICCOLO.}
    \centering
    \setlength{\tabcolsep}{8pt}
    \begin{tabular}{lrrrrr}
        \toprule
        Dataset & Split & POS & NEG & Total & Neg (\%) \\
        \midrule
        \multirow{1.75}{*}{REAL-Colon~\cite{realColon}}   & Train & 230{,}935 & 1{,}475{,}700 & 1{,}706{,}635 & 86.47 \\
        \multirow{1.75}{*}{(CC BY 4.0)}  & Val   & 29{,}067  &   330{,}236   &   359{,}303   & 91.91 \\
                     & Test  & 82{,}106  &   489{,}979   &   572{,}085   & 85.65 \\
        \midrule
        \multirow{1.75}{*}{SUN~\cite{sun}}  & Train & 34{,}915  &    58{,}839   &    93{,}754   & 62.76 \\
        \multirow{1.75}{*}{(Research Agreement)}     & Val   & 5{,}792   &    23{,}845   &    29{,}637   & 80.46 \\
                     & Test  & 8{,}429   &    26{,}869   &    35{,}298   & 76.12 \\
        \midrule
        \multirow{1.75}{*}{PICCOLO~\cite{piccolo}}      & Train & 2{,}183   &        20     &     2{,}203   &  0.91 \\
        \multirow{1.75}{*}{(CC BY 4.0)}   & Val   & 869       &        28     &       897     &  3.12 \\
                     & Test  & 332       &         1     &       333     &  0.30 \\
        \bottomrule
    \end{tabular}
    \label{tab:datasets}
\end{table}

Importantly, SUN and PICCOLO lack continuous, procedure-level recordings with consistent per-lesion identifiers. Consequently, lesion-level temporal metrics (e.g., persistence and detection latency) are only computable on REAL-Colon; SUN additionally supports frame-level alert metrics, whereas PICCOLO is limited to instance-level localization on still images. While SUN includes negative clips, it remains curated at the clip level and does not reflect the natural prevalence and temporal clustering of negative frames and procedure-related artifacts observed in full examinations.
\subsection{Curated In-Domain Baselines}
We first establish reference baselines on SUN and PICCOLO (\tablename~\ref{tab:curated_indomain_combined}) to mirror the standard evaluation settings prevalent in literature. The convolutional detectors achieve high detection-level accuracy under standard train/test alignment, whereas RT-DETR underperforms on the small PICCOLO training set (about $2{,}000$ images; $\mathrm{mAP}_{50}=0.321$ vs.\ $0.770$ for YOLOv11), as expected for the more data-hungry Transformer detectors. However, because PICCOLO contains an extremely small number of negative images in the official test split, it does not support a stable estimation of false-alarm rates, illustrating how highly curated benchmarks can overestimate deployment-relevant reliability.
\subsection{Transfer Asymmetry}
To probe the failure mode behind limited real-world deployment, we evaluate the cross-distribution transfer settings defined in Section~\ref{subsec:cross_distribution_transfer} (curated $\leftrightarrow$ procedure), without target-domain fine-tuning. For curated $\rightarrow$ procedure transfer, we report results under a matched false-alert operating point, selecting per-model thresholds such that the evaluated frame-level FPR is approximately comparable across models.

\begin{table*}[h]
    \centering
    \caption{Curated in-domain baselines on SUN and PICCOLO. Frame-level TPR/FPR are reported for SUN only; PICCOLO frame-level metrics are omitted because the official test split contains only one negative image. The convolutional detectors achieve high accuracy on both curated domains, while RT-DETR underperforms on PICCOLO.}
    \label{tab:curated_indomain_combined}
    \setlength{\tabcolsep}{4.7pt}
    \begin{tabular}{llccccc}
        \toprule
        Dataset (Train$\rightarrow$Test) & Model & $\tau^\star$ & mAP$_{50:95}$ & mAP$_{50}$ & TPR & FPR \\
        \midrule
        \multirow{4}{*}{SUN$\rightarrow$SUN}
            & Faster R-CNN & 0.07 & 0.408 & 0.738 & 0.739 & 0.044 \\
            & YOLOv8       & 0.07 & 0.396 & 0.694 & 0.731 & 0.046 \\
            & YOLOv11      & 0.10 & 0.411 & 0.724 & 0.735 & 0.047 \\
            & RT-DETR      & 0.12 & 0.389 & 0.689 & 0.725 & 0.050 \\
        \midrule
        \multirow{4}{*}{PICCOLO$\rightarrow$PICCOLO}
            & Faster R-CNN & --   & 0.531 & 0.743 & -- & -- \\
            & YOLOv8       & --   & 0.487 & 0.673 & -- & -- \\
            & YOLOv11      & --   & 0.560 & 0.770 & -- & -- \\
            & RT-DETR      & --   & 0.186 & 0.321 & -- & -- \\
        \bottomrule
    \end{tabular}
\end{table*}

\tablename~\ref{tab:transfer_asymmetry} shows a consistent transfer failure. Curated-trained models degrade sharply on REAL-Colon in both localization and alert sensitivity: YOLOv11 drops from $\mathrm{mAP}_{50}=0.724$ on SUN (cf.\ \tablename~\ref{tab:curated_indomain_combined}) to $0.164$ on REAL-Colon, and RT-DETR drops from $0.689$ to $0.225$. Models trained on the highly curated PICCOLO images degrade even further (YOLOv11 $\mathrm{mAP}_{50}=0.059$), indicating that lesion-centric benchmarks do not provide sufficient exposure to non-polyp content for full procedures.
\begin{table*}[tb]
    \centering
    \caption{Cross-distribution transfer between curated benchmarks (SUN, PICCOLO) and full procedures (REAL-Colon). For curated $\rightarrow$ REAL-Colon, thresholds are chosen to approximately match the evaluated frame-level FPR; reverse transfer reports detection-level generalization back to curated domains.
    }
    \label{tab:transfer_asymmetry}
    \setlength{\tabcolsep}{3.3pt}
    \begin{tabular}{llccccc}
        \toprule
        Dataset (Train$\rightarrow$Test) & Model & $\tau^\star$ & mAP$_{50:95}$ & mAP$_{50}$ & TPR & FPR \\
        \midrule

        \rowcolor{gray!15} \multicolumn{7}{l}{\textit{Curated $\rightarrow$ Procedure (Test: REAL-Colon)}} \\

        \multirow{4}{*}{\: SUN$\rightarrow$REAL-Colon}
            & Faster R-CNN & 0.20 & 0.131 & 0.220 & 0.290 & 0.043 \\
            & YOLOv8       & 0.25 & 0.105 & 0.177 & 0.261 & 0.045 \\
            & YOLOv11      & 0.04 & 0.097 & 0.164 & 0.240 & 0.040 \\
            & RT-DETR      & 0.25 & 0.132 & 0.225 & 0.310 & 0.040 \\
        \midrule
        \multirow{4}{*}{\: PICCOLO$\rightarrow$REAL-Colon}
            & Faster R-CNN & 0.99 & 0.080 & 0.154 & 0.271 & 0.060 \\
            & YOLOv8       & 0.80 & 0.024 & 0.051 & 0.175 & 0.046 \\
            & YOLOv11      & 0.70 & 0.031 & 0.059 & 0.134 & 0.041 \\
            & RT-DETR      & 0.80 & 0.005 & 0.012 & 0.007 & 0.003 \\

        \midrule
        \rowcolor{gray!15} \multicolumn{7}{l}{\textit{Procedure $\rightarrow$ Curated (Train: REAL-Colon)}} \\
        \multirow{4}{*}{\: REAL-Colon$\rightarrow$SUN}
            & Faster R-CNN & 0.12  & 0.251 & 0.503 & 0.514 & 0.049 \\
            & YOLOv8       & 0.006 & 0.365 & 0.714 & 0.767 & 0.048 \\
            & YOLOv11      & 0.06  & 0.361 & 0.705 & 0.765 & 0.043 \\
            & RT-DETR      & 0.08  & 0.379 & 0.717 & 0.728 & 0.043 \\
        \midrule
        \multirow{4}{*}{\: REAL-Colon$\rightarrow$PICCOLO}
            & Faster R-CNN & --    & 0.379 & 0.645 & -- & -- \\
            & YOLOv8       & --    & 0.314 & 0.558 & -- & -- \\
            & YOLOv11      & --    & 0.304 & 0.554 & -- & -- \\
            & RT-DETR      & --    & 0.336 & 0.602 & -- & -- \\
        \bottomrule
    \end{tabular}
\end{table*}
Procedure training, conversely, improves robustness: YOLOv8, YOLOv11, and RT-DETR trained on full procedures (REAL-Colon) retain competitive detection-level performance when transferred back to SUN ($\mathrm{mAP}_{50}=0.714$, $0.705$, and $0.717$), matching or exceeding their curated in-domain accuracy.
This asymmetry demonstrates that full-procedure data provide a more informative training signal than curated clips by better capturing the variability and negative-frame prevalence of continuous clinical video.
\subsection{Deployment-Oriented Analysis on REAL-Colon}
\label{sec:deployment}
Having established the need for full-procedure training, we characterize deployment relevant behavior on REAL-Colon (\tablename~\ref{tab:detection}). At a fixed threshold ($\tau=0.2$), RT-DETR achieves the highest spatial accuracy ($\mathrm{mAP}_{50}=0.488$) and frame-level sensitivity ($0.720$). However, fixed thresholds confound model quality with score calibration, since architectures incur different false-alert burdens at the same $\tau$: at $\tau=0.2$, RT-DETR yields a higher false-positive rate ($0.082$) than the YOLO models ($\mathrm{FPR}\le 0.020$), limiting comparability across architectures.
To enable a controlled comparison, we instead evaluate each model at its own matched operating point ($\tau^\star$), chosen independently per architecture such that its frame-level FPR falls within $4$--$5\%$. This equalizes false-alert burden across models, isolating differences in sensitivity and latency from differences in score calibration. Under this constraint the sensitivity gap narrows, with the YOLO detectors raising TPR substantially. Furthermore, RT-DETR now attains the lowest mean time-to-first-detection ($38.3$ frames) and the highest persistence, with YOLOv8 close behind ($46.8$ frames), reflecting an accuracy-versus-compute trade-off given the YOLO detectors' markedly higher throughput. Nonetheless these lesion-level differences fall within seed variance across the $n=21$ test lesions and are best read as trends rather than rankings.
\begin{table}[ht]
\centering
\setlength{\tabcolsep}{4pt}
\caption{REAL-Colon test performance at a fixed ($\tau=0.2$) and a normalized ($\tau^\star$ targeting $\mathrm{FPR}\approx 4$--$5\%$, selected post-hoc on the test split for cross-model normalization) threshold. Bold marks the best mean per row; lesion-level differences (block C, $n=21$) fall within seed variance and are not claimed as significant.}
\label{tab:detection}

\begin{tabular}{ll|cccc}
\toprule
 & Metric & F. R-CNN & YOLOv8 & YOLOv11 & RT-DETR \\
\midrule
& $\text{mAP}_{50}$ ($\uparrow$)    
            & $0.324 {\scriptstyle \pm 0.03}$
            & $0.413 {\scriptstyle \pm 0.00}$ 
            & $0.384 {\scriptstyle \pm 0.01}$
            & $\mathbf{0.488 {\scriptstyle \pm 0.02}}$  \\
\multirow{-2}{*}{A: Det.} & $\text{mAP}_{50{:}95}$ ($\uparrow$)  
            & $0.196 {\scriptstyle \pm 0.02}$  
            & $0.272 {\scriptstyle \pm 0.01}$ 
            & $0.245 {\scriptstyle \pm 0.01}$
            & $\mathbf{0.327 {\scriptstyle \pm 0.01}}$ \\
\midrule
& TPR $\tau$ ($\uparrow$)  
            & $0.463 {\scriptstyle \pm 0.06}$
            & $0.508 {\scriptstyle \pm 0.02}$
            & $0.453 {\scriptstyle \pm 0.00}$
            & $\mathbf{0.720 {\scriptstyle \pm 0.04}}$ \\
& FPR $\tau$ ($\downarrow$)
            & $0.046 {\scriptstyle \pm 0.02}$
            & $0.020 {\scriptstyle \pm 0.00}$
            & $\mathbf{0.017 {\scriptstyle \pm 0.00}}$
            & $0.082 {\scriptstyle \pm 0.03}$ \\
\rowcolor{gray!15} \cellcolor{white} & TPR $\tau^\star$ ($\uparrow$)  
            & $0.463 {\scriptstyle \pm 0.07}$
            & $0.605 {\scriptstyle \pm 0.03}$
            & $0.573 {\scriptstyle \pm 0.01}$
            & $\mathbf{0.651 {\scriptstyle \pm 0.05}}$ \\
\rowcolor{gray!15} \cellcolor{white} \multirow{-4}{*}{B: Frame} & FPR $\tau^\star$ ($\downarrow$)
            & $0.046 {\scriptstyle \pm 0.02}$
            & $0.044 {\scriptstyle \pm 0.01}$
            & $0.043 {\scriptstyle \pm 0.00}$
            & $\mathbf{0.043 {\scriptstyle \pm 0.02}}$ \\
\midrule
& Det. $\ge$1 $\tau$ ($\uparrow$)  
            & $\mathbf{21}$ 
            & $\mathbf{21}$ 
            & $\mathbf{21}$ 
            & $\mathbf{21}$ \\
& Det. $\ge$50\% $\tau$ ($\uparrow$)  
            & $11.7$
            & $12.7$
            & $12.7$
            & $\mathbf{19.7}$ \\
& Latency $\tau$ ($\downarrow$)
            & $63.9 {\scriptstyle \pm 42.5}$
            & $82.1 {\scriptstyle \pm 30.6}$
            & $103.0 {\scriptstyle \pm 37.6}$
            & $\mathbf{12.5 {\scriptstyle \pm 7.2}}$ \\
\rowcolor{gray!15} \cellcolor{white} & Det. $\ge$1 frame $\tau^\star$ ($\uparrow$)  
            & $\mathbf{21}$ 
            & $\mathbf{21}$ 
            & $\mathbf{21}$ 
            & $\mathbf{21}$ \\
\rowcolor{gray!15} \cellcolor{white} & Det. $\ge$50\% $\tau^\star$ ($\uparrow$)  
            & $11.7$
            & $16.0$
            & $15.7$
            & $\mathbf{17.3}$ \\
\rowcolor{gray!15} \cellcolor{white} \multirow{-6}{*}{C: Lesion} & Latency $\tau^\star$ ($\downarrow$)
            & $63.9 {\scriptstyle \pm 42.5}$
            & $46.8 {\scriptstyle \pm 26.3}$
            & $57.3 {\scriptstyle \pm 26.3}$
            & $\mathbf{38.3 {\scriptstyle \pm 17.3}}$ \\
\bottomrule
\end{tabular}
\end{table}
\section{Discussion and Conclusion}
Our experiments confirm that evaluation on curated benchmarks alone can overestimate clinical readiness. While state-of-the-art detectors reach high localization accuracy on SUN and PICCOLO, performance drops sharply on continuous, artifact-heavy full procedures. We attribute this to a pronounced transfer asymmetry: models trained on curated clips generalize poorly to REAL-Colon, consistent with limited exposure to prolonged negative segments and procedure-related artifacts, whereas procedure-trained models learn robust suppression of non-polyp content and transfer back to curated benchmarks.
Crucially, deployment cannot be judged by localization accuracy alone. Because score calibration varies by architecture, a fixed confidence threshold assigns different false-alert burdens to different models and is therefore misleading. Comparing detectors at a matched false-alert operating point ($\tau^\star$) removes this confound: at equalized false-alert burden, RT-DETR attains the lowest mean latency and the highest persistence, with the YOLO detectors close behind at markedly higher throughput, reflecting an accuracy-versus-compute trade-off rather than a uniform advantage for either architecture family.
These findings motivate a concrete requirement: full-procedure video is needed for both training and evaluation when the goal is deployment, since false-alert burden, time-to-first-detection, and temporal reliability can only be measured under realistic negative-frame prevalence. As our detectors are memoryless and operate per frame, incorporating temporal modeling that exploits inter-frame consistency is a natural next step toward lower latency and steadier detections.
\subsubsection{Limitations.} Several factors bound these conclusions. First, the datasets differ not only in realism but also in scale, annotation protocol, and negative-frame prevalence, so we cannot attribute the transfer asymmetry to realism alone. The core of our argument is nonetheless scale-independent: lesion-centric benchmarks cannot measure false-alert burden, latency, or temporal reliability, regardless of their size. Second, the detectors use their default framework recipes (Detectron2 and Ultralytics) at different batch sizes, so cross-architecture comparisons reflect architectures as commonly deployed rather than a controlled architectural study. Third, $\tau^\star$ is selected on the evaluated split for normalization, and a deployment system should instead fix this operating point on a separate validation set. Fourth, our false-alert measure is frame-level and therefore weights every alerting negative frame equally: a single persistent alert spanning many frames and an equal number of isolated interruptions score identically, and false detections on lesion-bearing frames are not counted at all. Event-level rates, such as alerts per minute or per procedure, might map more directly onto the interruption burden a clinician actually experiences. Fifth, the curated-versus-procedure contrast may also confound negative-frame prevalence with domain difference; an in-domain control training on REAL-Colon's lesion-bearing frames alone would separate the two, and mitigation strategies such as hard-negative mining remain unexplored here. Finally, detection is dominated by medium and large lesions, with near-zero AP on small polyps, and our lesion-level evidence comes from a single cohort ($n=21$ test lesions). Replicating the asymmetry and operating-point behavior across additional cohorts, sites, and acquisition settings is the clear next step.

\begin{credits}
\subsubsection{\ackname} This study was funded through the Hightech Agenda Bayern (HTA) of the Free State of Bavaria, Germany.

\subsubsection{\discintname} The authors have no competing interests to declare that are relevant to the content of this article. 
\end{credits}
%
%
%
\bibliographystyle{splncs04}
\bibliography{mybibliography}
%






\clearpage
\appendix
\setcounter{page}{1}\renewcommand{\thepage}{S\arabic{page}}

\renewcommand{\theHsection}{supp.\thesection}
\renewcommand{\theHtable}{supp.\thetable}
\renewcommand{\theHfigure}{supp.\thefigure}
\renewcommand{\theHequation}{supp.\theequation}

\numberwithin{table}{section}
\renewcommand{\thetable}{\thesection\arabic{table}}

\numberwithin{figure}{section}
\renewcommand{\thefigure}{\thesection\arabic{figure}}

\numberwithin{equation}{section}
\renewcommand{\theequation}{\thesection\arabic{equation}}


\section*{Supplementary Material}
\label{sec:sup-overview}

This supplementary material complements the main paper with additional
experimental detail and analyses.
\textbf{Section~\ref{sec:sup-setup}} gives the full computational setup needed to reproduce the
reported statistics.
\textbf{Section~\ref{sec:sup-data}} releases the exact REAL-Colon per-video
frame ranges and the SUN case-level split.
\textbf{Section~\ref{sec:sup-results}} extends the REAL-Colon evaluation across
localization accuracy, operating-point behavior, temporal reliability,
runtime, and ablations, and confirms the main-paper ranking: RT-DETR
leads localization ($\mathrm{mAP}_{50}=0.488$) and detects lesions
earliest (first-frame latency $12.5$ frames), detection is carried
almost entirely by medium and large lesions (near-zero small-polyp AP),
and the open detectors reach commercial-CADe sensitivity
($0.60$--$0.72$).
\textbf{Section~\ref{sec:sup-metrics}} defines every detection-level,
frame-level, AFROC, and runtime metric used throughout.
\subsubsection{Data use declaration.}
This study uses three previously published datasets (REAL-Colon,
SUN, and PICCOLO) and cites the corresponding original
publications~\cite{realColon,sun,piccolo}.
These datasets were used for academic research and benchmarking in
accordance with the terms and usage conditions provided by their
publishers.
Any ethics approvals and consent procedures related to data collection
are those reported in the respective original publications.
\section{Computational Setup}
\label{sec:sup-setup}
\setcounter{table}{0}\setcounter{figure}{0}
To make every reported statistic reproducible, all experiments were run
on two fixed workstations with pinned drivers and dependencies.
Workstation~1 is the primary environment for all REAL-Colon experiments
and final evaluations. Workstation~2 is used only for training on the SUN and PICCOLO datasets.

\subsection{Hardware}
\subsubsection{Workstation~1 (primary):}
Two Intel~Xeon~Gold~5416S CPUs (64 threads total), 503\,GiB system
memory, two NVIDIA~H100~NVL GPUs (95\,830\,MiB VRAM each), with data
stored on a local NVMe~SSD.

\subsubsection{Workstation~2 (supplementary training):}
Intel~Xeon~W-2265 (24 threads), 125\,GiB system memory, two
NVIDIA~RTX~A5000 GPUs (24\,564\,MiB VRAM each).

\subsection{Software}
Both systems run Ubuntu~24.04.3~LTS.
CUDA~12.9 / driver~575.57.08 (WS1) and driver~580.95.05 (WS2).
Two Conda environments were maintained:
\begin{itemize}
  \item \textbf{Ultralytics} (shared): Python~3.10.19,
        \texttt{ultralytics}~8.3.232, PyTorch~2.6.0+cu124.
        Used for all YOLOv8, YOLOv11, and RT-DETR runs~\cite{rtdetr_ultralytics}.
  \item \textbf{Detectron2} (WS1 only): Python~3.10.19,
        \texttt{pycocotools}~2.0.10, Detectron2~0.6,
        PyTorch~2.5.1+cu121.
        Used for all Faster~R-CNN runs~\cite{detectron2}.
\end{itemize}
\newpage
\section{Dataset Partitioning Details}
\label{sec:sup-data}
\setcounter{table}{0}\setcounter{figure}{0}
To make the partition behind Table~\ref{tab:datasets} of the main paper
fully reproducible, we release the exact frame- and case-level
assignments for REAL-Colon and SUN.

\subsection{REAL-Colon: Per-Video Frame Ranges}
\label{sec:sup-data-realcolon}
Colonoscopy recordings include out-of-patient segments (insertion and
withdrawal) that carry no diagnostic content. We therefore retain, for
each of the 60 REAL-Colon~\cite{realColon} videos, only the
inside-patient frame-ID interval of Table~\ref{tab:sup-frame-ranges}.
\begin{table}[htbp]
\centering
\caption{Retained inside-patient frame-ID intervals for the 60 REAL-Colon
         videos (labeled patient\_video). Frames outside the listed range
         are discarded as out-of-patient.}
\label{tab:sup-frame-ranges}
\small
\setlength{\tabcolsep}{6pt}
\renewcommand{\arraystretch}{1.15}
\begin{tabular}{llllll}
\toprule
\textbf{Video} & \textbf{Range} &
\textbf{Video} & \textbf{Range} &
\textbf{Video} & \textbf{Range} \\
\midrule
1\_1  & 920--45137   & 1\_2  & 1620--25974  & 1\_3  & 1967--39372 \\
1\_4  & 850--45385   & 1\_5  & 360--28544   & 1\_6  & 786--48217 \\
1\_7  & 1050--37920  & 1\_8  & 1165--34891  & 1\_9  & 2031--39884 \\
1\_10 & 2075--46215  & 1\_11 & 380--23390   & 1\_12 & 530--40842 \\
1\_13 & 1996--22417  & 1\_14 & 347--28711   & 1\_15 & 910--30831 \\
\midrule
2\_1  & 1--25421     & 2\_2  & 414--24916   & 2\_3  & 1--47191 \\
2\_4  & 1--30537     & 2\_5  & 70--40342    & 2\_6  & 670--52180 \\
2\_7  & 989--49622   & 2\_8  & 568--32638   & 2\_9  & 1447--28957 \\
2\_10 & 312--25540   & 2\_11 & 420--37747   & 2\_12 & 1--33739 \\
2\_13 & 330--22515   & 2\_14 & 785--55308   & 2\_15 & 583--20772 \\
\midrule
3\_1  & 415--51456   & 3\_2  & 990--49874   & 3\_3  & 12240--105451 \\
3\_4  & 6342--75787  & 3\_5  & 4980--122040 & 3\_6  & 4478--102060 \\
3\_7  & 6932--58000  & 3\_8  & 1702--50387  & 3\_9  & 1071--58101 \\
3\_10 & 452--55736   & 3\_11 & 1617--62313  & 3\_12 & 1041--96306 \\
3\_13 & 773--73379   & 3\_14 & 2427--75752  & 3\_15 & 2757--110858 \\
\midrule
4\_1  & 699--25264   & 4\_2  & 687--34245   & 4\_3  & 1491--33222 \\
4\_4  & 583--33629   & 4\_5  & 298--21781   & 4\_6  & 326--40377 \\
4\_7  & 593--28828   & 4\_8  & 1145--22604  & 4\_9  & 1600--30189 \\
4\_10 & 925--23927   & 4\_11 & 786--48100   & 4\_12 & 1338--22971 \\
4\_13 & 1300--54128  & 4\_14 & 459--36698   & 4\_15 & 327--53698 \\
\bottomrule
\end{tabular}
\end{table}

\subsection{SUN: Case-Level Split Assignment}
\label{sec:sup-data-sun}
To prevent patient-level leakage, SUN is
partitioned at the case level rather than by frame.
Table~\ref{tab:sup-sun-split} gives the resulting per-case image counts
and train/val/test assignment for all 113 SUN~\cite{sun} cases
(positive IDs~1--100, negative IDs~101--113).

\begin{table}[htbp]
\centering
\caption{SUN case-level image counts and train/val/test assignment
         (positive cases IDs~1--100, negative cases~101--113).
         Splitting by case rather than by frame prevents patient-level
         leakage into the evaluation folds.}
\label{tab:sup-sun-split}
\scriptsize
\setlength{\tabcolsep}{5pt}
\renewcommand{\arraystretch}{1.1}
\begin{tabular}{cccccccc}
\toprule
\multicolumn{4}{c}{Cases 1--57} & \multicolumn{4}{c}{Cases 58--113} \\
\cmidrule(lr){1-4}\cmidrule(lr){5-8}
Case ID & Type & Split & \#Images &
Case ID & Type & Split & \#Images \\
\midrule
1  & POS & train & 527  & 58  & POS & train & 267   \\
2  & POS & train & 1313 & 59  & POS & train & 646   \\
3  & POS & train & 292  & 60  & POS & train & 146   \\
4  & POS & train & 80   & 61  & POS & train & 679   \\
5  & POS & train & 930  & 62  & POS & train & 351   \\
6  & POS & train & 491  & 63  & POS & train & 632   \\
7  & POS & train & 315  & 64  & POS & train & 81    \\
8  & POS & train & 256  & 65  & POS & train & 222   \\
9  & POS & train & 136  & 66  & POS & train & 1685  \\
10 & POS & train & 436  & 67  & POS & train & 191   \\
11 & POS & train & 113  & 68  & POS & train & 1319  \\
12 & POS & train & 538  & 69  & POS & train & 130   \\
13 & POS & train & 479  & 70  & POS & train & 264   \\
14 & POS & train & 1183 & 71  & POS & val   & 1021  \\
15 & POS & train & 487  & 72  & POS & val   & 774   \\
16 & POS & train & 199  & 73  & POS & val   & 1285  \\
17 & POS & train & 304  & 74  & POS & val   & 276   \\
18 & POS & train & 243  & 75  & POS & val   & 343   \\
19 & POS & train & 96   & 76  & POS & val   & 343   \\
20 & POS & train & 3159 & 77  & POS & val   & 215   \\
21 & POS & train & 100  & 78  & POS & val   & 267   \\
22 & POS & train & 314  & 79  & POS & val   & 76    \\
23 & POS & train & 182  & 80  & POS & val   & 1192  \\
24 & POS & train & 973  & 81  & POS & test  & 427   \\
25 & POS & train & 338  & 82  & POS & test  & 111   \\
26 & POS & train & 370  & 83  & POS & test  & 795   \\
27 & POS & train & 249  & 84  & POS & test  & 218   \\
28 & POS & train & 195  & 85  & POS & test  & 1393  \\
29 & POS & train & 377  & 86  & POS & test  & 257   \\
30 & POS & train & 224  & 87  & POS & test  & 454   \\
31 & POS & train & 183  & 88  & POS & test  & 249   \\
32 & POS & train & 981  & 89  & POS & test  & 149   \\
33 & POS & train & 594  & 90  & POS & test  & 479   \\
34 & POS & train & 245  & 91  & POS & test  & 1061  \\
35 & POS & train & 1212 & 92  & POS & test  & 391   \\
36 & POS & train & 815  & 93  & POS & test  & 452   \\
37 & POS & train & 448  & 94  & POS & test  & 136   \\
38 & POS & train & 509  & 95  & POS & test  & 606   \\
39 & POS & train & 713  & 96  & POS & test  & 301   \\
40 & POS & train & 159  & 97  & POS & test  & 431   \\
41 & POS & train & 108  & 98  & POS & test  & 170   \\
42 & POS & train & 268  & 99  & POS & test  & 161   \\
43 & POS & train & 260  & 100 & POS & test  & 188   \\
44 & POS & train & 745  & 101 & NEG & train & 9960  \\
45 & POS & train & 383  & 102 & NEG & test  & 10073 \\
46 & POS & train & 170  & 103 & NEG & train & 7152  \\
47 & POS & train & 705  & 104 & NEG & train & 14635 \\
48 & POS & train & 176  & 105 & NEG & train & 7916  \\
49 & POS & train & 181  & 106 & NEG & val   & 17046 \\
50 & POS & train & 740  & 107 & NEG & test  & 5636  \\
51 & POS & train & 1737 & 108 & NEG & train & 2568  \\
52 & POS & train & 207  & 109 & NEG & train & 9522  \\
53 & POS & train & 245  & 110 & NEG & train & 7086  \\
54 & POS & train & 345  & 111 & NEG & test  & 4832  \\
55 & POS & train & 700  & 112 & NEG & val   & 6799  \\
56 & POS & train & 248  & 113 & NEG & test  & 6328  \\
57 & POS & train & 326  &     &     &       &       \\
\midrule
\multicolumn{3}{l}{Total positive images} & 49{,}136 &
\multicolumn{3}{l}{Total negative images} & 109{,}553 \\
\multicolumn{3}{l}{Total images} & 158{,}689 & & & & \\
\bottomrule
\end{tabular}
\end{table}

\newpage
\section{Extended Experimental Results}
\label{sec:sup-results}
\setcounter{table}{0}\setcounter{figure}{0}
\subsection{Detection Accuracy on REAL-Colon}
\label{sec:sup-det-accuracy}
The main-paper REAL-Colon results (Table~\ref{tab:detection}) report
$\mathrm{mAP}_{50}$ and $\mathrm{mAP}_{50:95}$. To characterize
localization more finely and separate ranking quality from recall
saturation, we report the stricter $\mathrm{mAP}_{75}$ and average
recall (AR) at 1, 10, and 100 detections per image, and break average
precision down by object size following the COCO definition
(small ${\le}32^2$, medium $32^2$--$96^2$, large ${\ge}96^2$
pixels)~\cite{coco}. All detectors use the three-seed protocol of the
main experiments.

\subsubsection{Results}
Across every localization threshold and recall budget, RT-DETR is the
strongest detector (Table~\ref{tab:sup-det-full}): it reaches
$\mathrm{mAP}_{75}=0.359$ and $\mathrm{AR}_{100}=0.672$, ahead of the best
convolutional detector (YOLOv8, $\mathrm{mAP}_{75}=0.299$,
$\mathrm{AR}_{100}=0.471$). The margin widens as the recall budget grows
(RT-DETR $\mathrm{AR}_{1}{\to}\mathrm{AR}_{100}$: $0.415{\to}0.672$),
showing that its advantage comes from recovering additional lesions
rather than from ranking alone. By object size
(Table~\ref{tab:sup-det-size}), average precision is carried by medium
and large lesions: small-polyp AP is at or near zero for every
architecture (${\le}0.002$), so detection in this domain is effectively a
medium- and large-lesion problem.
\begin{table}[htbp]
    \centering
    \caption{Extended detection-level performance on REAL-Colon
         (mean\,$\pm$\,std over three seeds).
         $\mathrm{mAP}_{t}$ denotes mean average precision at IoU
         threshold~$t$, $\mathrm{mAP}_{50:95}$ averages over IoU
         $0.50$--$0.95$, and $\mathrm{AR}_{k}$ is average recall allowing
         $k$ detections per image.
         RT-DETR leads at every threshold and recall budget.}
    \label{tab:sup-det-full}
    \setlength{\tabcolsep}{5pt}
    \renewcommand{\arraystretch}{1.15}
    \resizebox{\linewidth}{!}{%
    \begin{tabular}{lcccc}
        \toprule
          & Faster R-CNN & YOLOv8 & YOLOv11 & RT-DETR \\
        \midrule
        $\mathrm{mAP}_{50}$    
            & $0.324 {\scriptstyle \pm 0.032}$ 
            & $0.413 {\scriptstyle \pm 0.002}$ 
            & $0.384 {\scriptstyle \pm 0.009}$
            & $0.488 {\scriptstyle \pm 0.019}$  \\
        $\mathrm{mAP}_{75}$     
            & $0.216 {\scriptstyle \pm 0.023}$ 
            & $0.299 {\scriptstyle \pm 0.010}$ 
            & $0.263 {\scriptstyle \pm 0.005}$
            & $0.359 {\scriptstyle \pm 0.009}$ \\
        $\mathrm{mAP}_{50:95}$ 
            & $0.196 {\scriptstyle \pm 0.021}$  
            & $0.272 {\scriptstyle \pm 0.005}$ 
            & $0.245 {\scriptstyle \pm 0.005}$
            & $0.327 {\scriptstyle \pm 0.009}$ \\
        $\mathrm{AR}_{1}$      
            & $0.295 {\scriptstyle \pm 0.014}$  
            & $0.376 {\scriptstyle \pm 0.003}$ 
            & $0.357 {\scriptstyle \pm 0.005}$
            & $0.415 {\scriptstyle \pm 0.011}$ \\
        $\mathrm{AR}_{10}$     
            & $0.387 {\scriptstyle \pm 0.030}$  
            & $0.466 {\scriptstyle \pm 0.009}$ 
            & $0.453 {\scriptstyle \pm 0.012}$
            & $0.585 {\scriptstyle \pm 0.007}$ \\
        $\mathrm{AR}_{100}$    
            & $0.394 {\scriptstyle \pm 0.035}$  
            & $0.471 {\scriptstyle \pm 0.012}$ 
            & $0.457 {\scriptstyle \pm 0.013}$
            & $0.672 {\scriptstyle \pm 0.008}$ \\
        \bottomrule
    \end{tabular}%
    }
\end{table}

\begin{table}[htbp]
\centering
\caption{Average precision by object size on REAL-Colon, following the
         COCO definition (small ${\le}32^2$, medium $32^2$--$96^2$, large
         ${\ge}96^2$ pixels), as mean\,$\pm$\,std over three seeds.
         Small-polyp AP is near zero for all detectors, so localization is
         driven by medium and large lesions.}
\label{tab:sup-det-size}
\setlength{\tabcolsep}{5pt}
\renewcommand{\arraystretch}{1.15}
\resizebox{\linewidth}{!}{%
\begin{tabular}{lcccc}
\toprule
 & \textbf{Faster R-CNN} & \textbf{YOLOv8} & \textbf{YOLOv11} & \textbf{RT-DETR} \\
\midrule
$\mathrm{AP}_{\text{small}}$
    & $0.000$ & $0.000$ & $0.000$ & $0.002$ \\
$\mathrm{AP}_{\text{medium}}$
    & $0.012 {\scriptstyle \pm 0.006}$ & $0.049 {\scriptstyle \pm 0.009}$ & $0.054 {\scriptstyle \pm 0.005}$ & $0.044 {\scriptstyle \pm 0.014}$ \\
$\mathrm{AP}_{\text{large}}$
    & $0.207 {\scriptstyle \pm 0.021}$ & $0.284 {\scriptstyle \pm 0.005}$ & $0.255 {\scriptstyle \pm 0.005}$ & $0.343 {\scriptstyle \pm 0.010}$ \\
\bottomrule
\end{tabular}%
}
\end{table}

\subsection{AFROC Operating Curves}
\label{sec:sup-afroc}
\begin{figure}[htbp]
\centering
\includegraphics[width=0.495\linewidth]{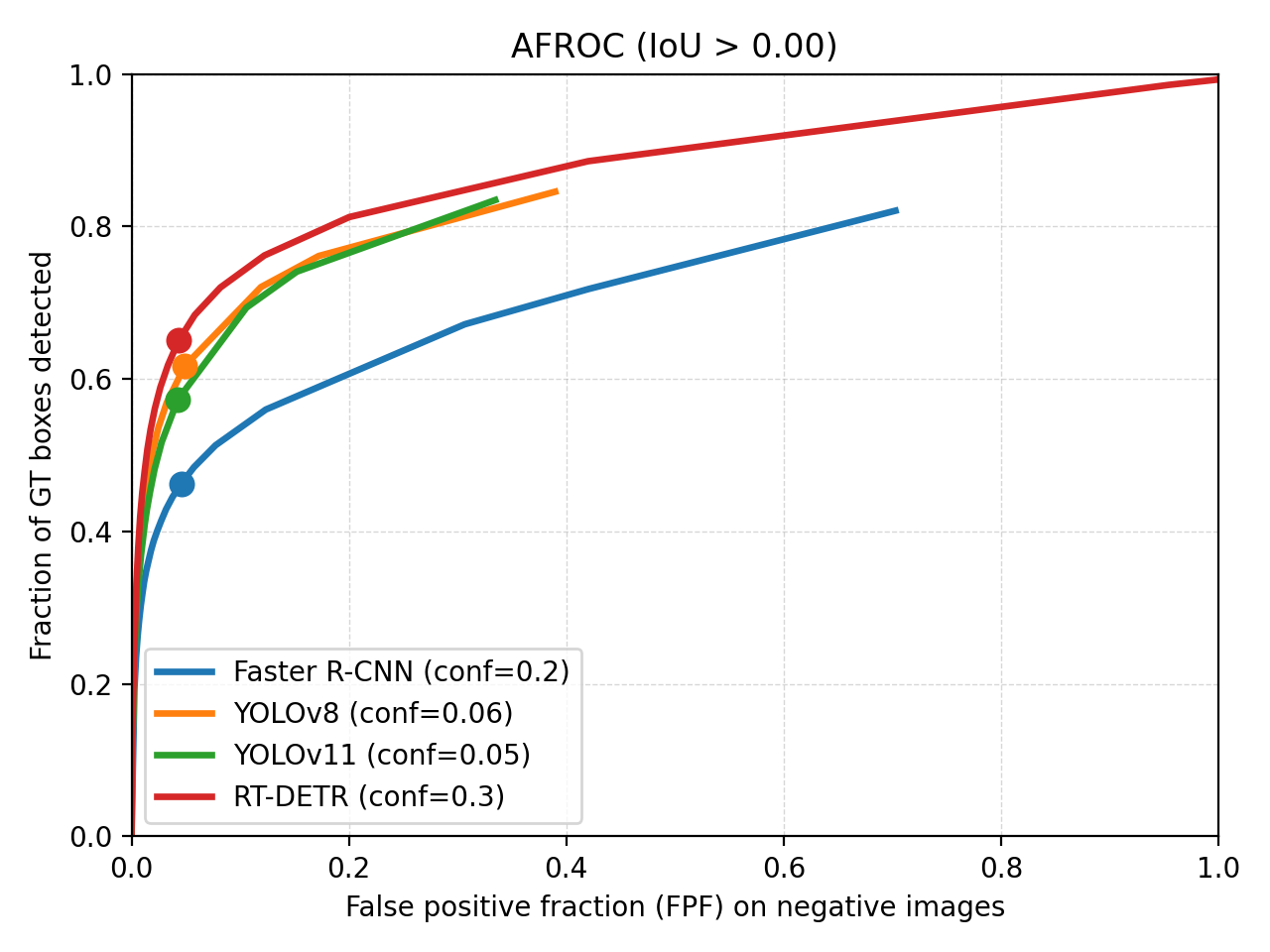}
\includegraphics[width=0.495\linewidth]{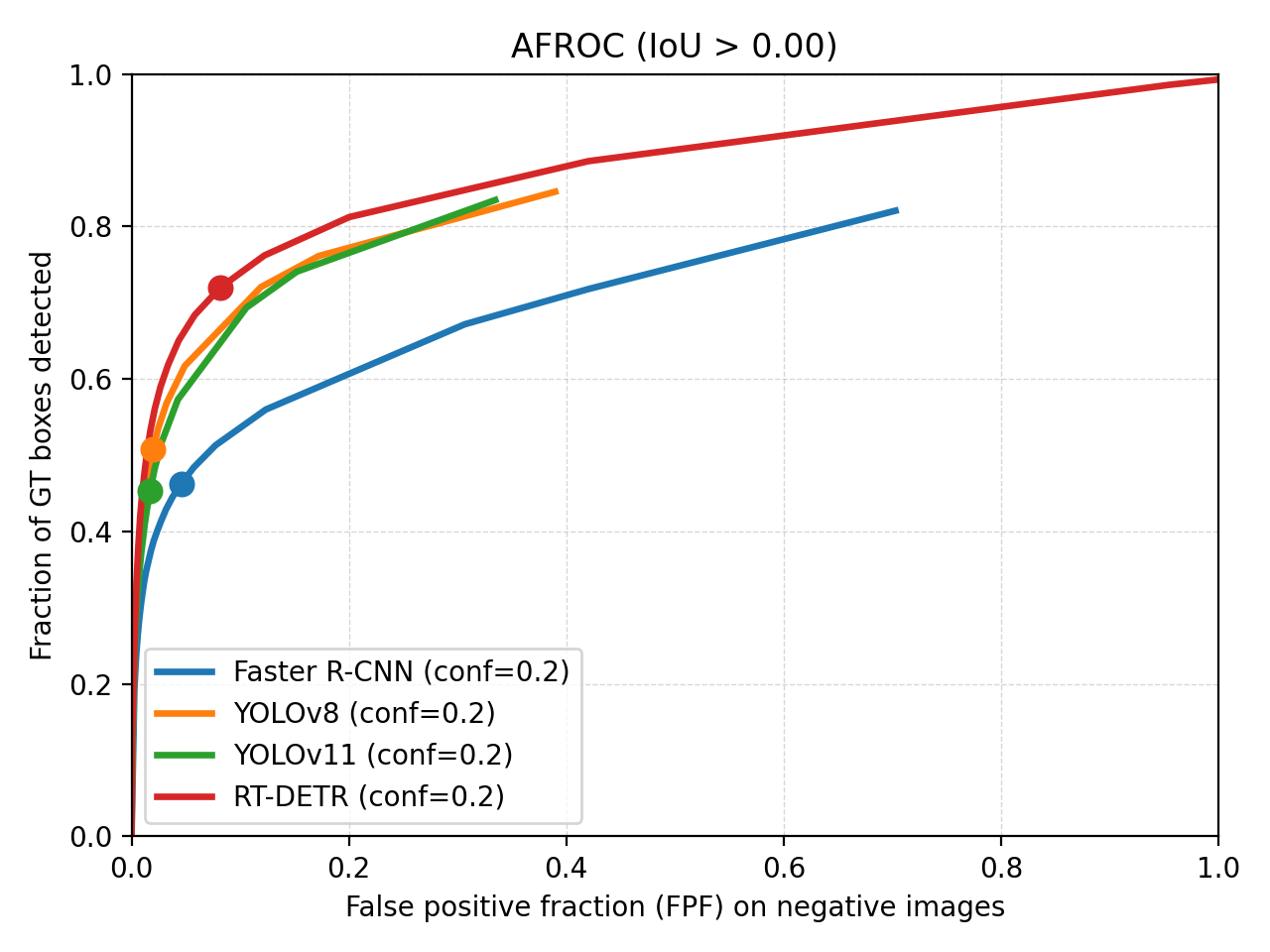}
\caption{AFROC curves on REAL-Colon.
         \textbf{Left:} model-specific confidence thresholds
         ($\mathrm{FPR}\approx4$--$5\%$).
         \textbf{Right:} fixed $\tau=0.2$ for all models.
         The $y$-axis is the fraction of ground-truth lesion boxes
         detected and the $x$-axis is the fraction of negative frames
         raising at least one false-positive alert.
         RT-DETR retains the highest sensitivity at the lowest
         false-positive fractions.}
\label{fig:sup-afroc}
\end{figure}

A single operating point cannot show how sensitivity trades against the
false-alert burden across the full threshold range. We therefore report
Alternative Free-response ROC (AFROC) curves on the REAL-Colon test set,
which plot lesion-level sensitivity against the false-positive fraction
(FPF, defined in Section~\ref{sec:sup-metrics-afroc}) and so extend the
fixed-threshold frame-level comparison in Table~\ref{tab:detection}.
Figure~\ref{fig:sup-afroc} shows two panels: the left evaluates each model
at its own operating threshold (matched to $\mathrm{FPR}\approx4$--$5\%$),
and the right evaluates all models at a shared $\tau=0.2$ for a direct
comparison.

\subsubsection{Results}
RT-DETR dominates at low false-positive fractions, confirming its
operating-point advantage from Table~\ref{tab:detection}: it retains the
highest sensitivity where false alerts are rarest. As the FPF budget
grows, the YOLO detectors approach its sensitivity, so the ranking is
widest in the low-false-alert regime that matters most clinically.

\subsection{Per-Seed Confidence Threshold Analysis}
\label{sec:sup-per-seed}
The main-paper operating points (Table~\ref{tab:detection}) use a fixed,
model-specific confidence threshold per detector. To show that this
choice is stable and to justify it over per-split tuning, we compare
three threshold-selection strategies: (i)~a shared $\tau=0.2$ for all
models, (ii)~model-specific thresholds tuned once to
$\mathrm{FPR}\approx4$--$5\%$ (Faster\,R-CNN: $0.20$; YOLOv8: $0.06$;
YOLOv11: $0.05$; RT-DETR: $0.30$), and (iii)~thresholds refitted
independently per random seed (Faster\,R-CNN: $0.10/0.20/0.26$; RT-DETR:
$0.32/0.20/0.32$; YOLO unchanged across seeds).
Table~\ref{tab:sup-det-conf} reports detection-level precision and recall
under all three, and Table~\ref{tab:sup-frame-conf} extends the
frame-level block of Table~\ref{tab:detection} with the full metric set
(specificity, precision, $F_1$, $F_2$) under the model-specific and
per-seed strategies.

\subsubsection{Results}
A single shared threshold is the wrong comparison: at $\tau=0.2$ RT-DETR
over-fires and its precision falls to $0.398$ while recall reaches
$0.720$, whereas at its own threshold precision recovers to $0.577$ at
$0.651$ recall (Table~\ref{tab:sup-det-conf}). Once each detector sits at
its model-specific operating point, the ranking matches the main paper:
RT-DETR attains the best frame-level $F_1$ ($0.681$) and $F_2$ ($0.662$),
ahead of YOLOv8 ($F_1=0.648$, $F_2=0.621$)
(Table~\ref{tab:sup-frame-conf}). Per-seed refitting lowers cross-seed
variance but changes the means only marginally, and we do not recommend
it for deployment: thresholds fitted to the evaluated split introduce
optimistic selection bias and are impractical where a fixed operating
point must be set in advance.
\begin{table}[htbp]
\centering
\caption{Detection-level precision and recall under three
         confidence-threshold strategies (shared $\tau=0.2$,
         model-specific, and per-seed; mean\,$\pm$\,std over three seeds).
         A shared threshold penalizes RT-DETR, which over-fires at low
         $\tau$; model-specific thresholds place every detector at a
         comparable false-positive rate.}
\label{tab:sup-det-conf}
\setlength{\tabcolsep}{5pt}
\renewcommand{\arraystretch}{1.15}
\resizebox{\linewidth}{!}{%
\begin{tabular}{lcccc}
\toprule
\textbf{Metric} &
\textbf{Faster R-CNN} & \textbf{YOLOv8} &
\textbf{YOLOv11}      & \textbf{RT-DETR} \\
\midrule
\multicolumn{5}{l}{\emph{Shared confidence $\tau=0.20$}} \\
\midrule
Precision & $0.524 {\scriptstyle \pm 0.086}$ & $0.701 {\scriptstyle \pm 0.022}$ & $0.715 {\scriptstyle \pm 0.011}$ & $0.398 {\scriptstyle \pm0.095}$ \\
Recall    & $0.463 {\scriptstyle \pm 0.066}$ & $0.508 {\scriptstyle \pm 0.022}$ & $0.453 {\scriptstyle \pm 0.004}$ & $0.720 {\scriptstyle \pm 0.041}$ \\
\midrule
\multicolumn{5}{l}{\emph{Model-specific thresholds (FPR\,$\approx$\,4--5\%)}} \\
\midrule
Precision & $0.524 {\scriptstyle \pm 0.086}$ & $0.523 {\scriptstyle \pm 0.029}$ & $0.516 {\scriptstyle \pm 0.017}$ & $0.577 {\scriptstyle \pm 0.094}$ \\
Recall    & $0.463 {\scriptstyle \pm 0.066}$ & $0.605 {\scriptstyle \pm 0.030}$ & $0.573 {\scriptstyle \pm 0.009}$ & $0.651 {\scriptstyle \pm 0.049}$ \\
\midrule
\multicolumn{5}{l}{\emph{Per-seed thresholds}} \\
\midrule
Precision & $0.505 {\scriptstyle \pm 0.012}$ & $0.523 {\scriptstyle \pm 0.029}$ & $0.516 {\scriptstyle \pm 0.017}$ & $0.537 {\scriptstyle \pm 0.027}$ \\
Recall    & $0.470 {\scriptstyle \pm 0.055}$ & $0.605 {\scriptstyle \pm 0.030}$ & $0.573 {\scriptstyle \pm 0.009}$ & $0.669 {\scriptstyle \pm 0.006}$ \\
\bottomrule
\end{tabular}%
}
\end{table}

\begin{table}[htbp]
\centering
\caption{Frame-level performance under the model-specific and per-seed
         threshold strategies (mean\,$\pm$\,std over three seeds),
         extending the frame-level block of Table~\ref{tab:detection} with
         specificity, precision, $F_1$, and $F_2$.
         At matched operating points RT-DETR attains the best $F_1$ and
         $F_2$. Per-seed refitting only reduces variance.}
\label{tab:sup-frame-conf}
\setlength{\tabcolsep}{5pt}
\renewcommand{\arraystretch}{1.15}
\resizebox{\linewidth}{!}{%
\begin{tabular}{lcccc}
\toprule
\textbf{Metric} &
\textbf{Faster R-CNN} & \textbf{YOLOv8} &
\textbf{YOLOv11}      & \textbf{RT-DETR} \\
\midrule
\multicolumn{5}{l}{\emph{Model-specific thresholds (FPR\,$\approx$\,4--5\%)}} \\
\midrule
Sensitivity / TPR & $0.463 {\scriptstyle \pm 0.066}$ & $0.605 {\scriptstyle \pm 0.030}$ & $0.573 {\scriptstyle \pm 0.009}$ & $0.651 {\scriptstyle \pm 0.049}$ \\
Specificity       & $0.954 {\scriptstyle \pm 0.017}$ & $0.956 {\scriptstyle \pm 0.006}$ & $0.957 {\scriptstyle \pm 0.003}$ & $0.957 {\scriptstyle \pm 0.017}$ \\
FPR               & $0.046 {\scriptstyle \pm 0.017}$ & $0.044 {\scriptstyle \pm 0.006}$ & $0.043 {\scriptstyle \pm 0.003}$ & $0.043 {\scriptstyle \pm 0.017}$ \\
Precision         & $0.637 {\scriptstyle \pm 0.086}$ & $0.698 {\scriptstyle \pm 0.023}$ & $0.693 {\scriptstyle \pm 0.020}$ & $0.723 {\scriptstyle \pm 0.070}$ \\
$F_1$ score       & $0.532 {\scriptstyle \pm0 .047}$ & $0.648 {\scriptstyle \pm 0.015}$ & $0.627 {\scriptstyle \pm 0.013}$ & $0.681 {\scriptstyle \pm 0.001}$ \\
$F_2$ score       & $0.488 {\scriptstyle \pm 0.057}$ & $0.621 {\scriptstyle \pm 0.024}$ & $0.594 {\scriptstyle \pm 0.011}$ & $0.662 {\scriptstyle \pm 0.031}$ \\
\midrule
\multicolumn{5}{l}{\emph{Per-seed thresholds}} \\
\midrule
Sensitivity / TPR & $0.470 {\scriptstyle \pm 0.055}$ & $0.605 {\scriptstyle \pm 0.030}$ & $0.573 {\scriptstyle \pm 0.009}$ & $0.669 {\scriptstyle \pm 0.006}$ \\
Specificity       & $0.953 {\scriptstyle \pm 0.003}$ & $0.956 {\scriptstyle \pm 0.006}$ & $0.957 {\scriptstyle \pm 0.003}$ & $0.952 {\scriptstyle \pm 0.001}$ \\
FPR               & $0.047 {\scriptstyle \pm 0.003}$ & $0.044 {\scriptstyle \pm 0.006}$ & $0.043 {\scriptstyle \pm 0.003}$ & $0.048 {\scriptstyle \pm 0.001}$ \\
Precision         & $0.625 {\scriptstyle \pm 0.039}$ & $0.698 {\scriptstyle \pm 0.023}$ & $0.693 {\scriptstyle \pm 0.020}$ & $0.702 {\scriptstyle \pm 0.005}$ \\
$F_1$ score       & $0.536 {\scriptstyle \pm 0.050}$ & $0.648 {\scriptstyle \pm 0.015}$ & $0.627 {\scriptstyle \pm 0.013}$ & $0.685 {\scriptstyle \pm 0.003}$ \\
$F_2$ score       & $0.494 {\scriptstyle \pm 0.054}$ & $0.621 {\scriptstyle \pm 0.024}$ & $0.594 {\scriptstyle \pm 0.011}$ & $0.675 {\scriptstyle \pm 0.005}$ \\
\bottomrule
\end{tabular}%
}
\end{table}

\subsection{Lesion-Level Consistency and Early Detection}
\label{sec:sup-lesion}
\begin{figure}[p]
\centering
\includegraphics[width=0.87\linewidth]{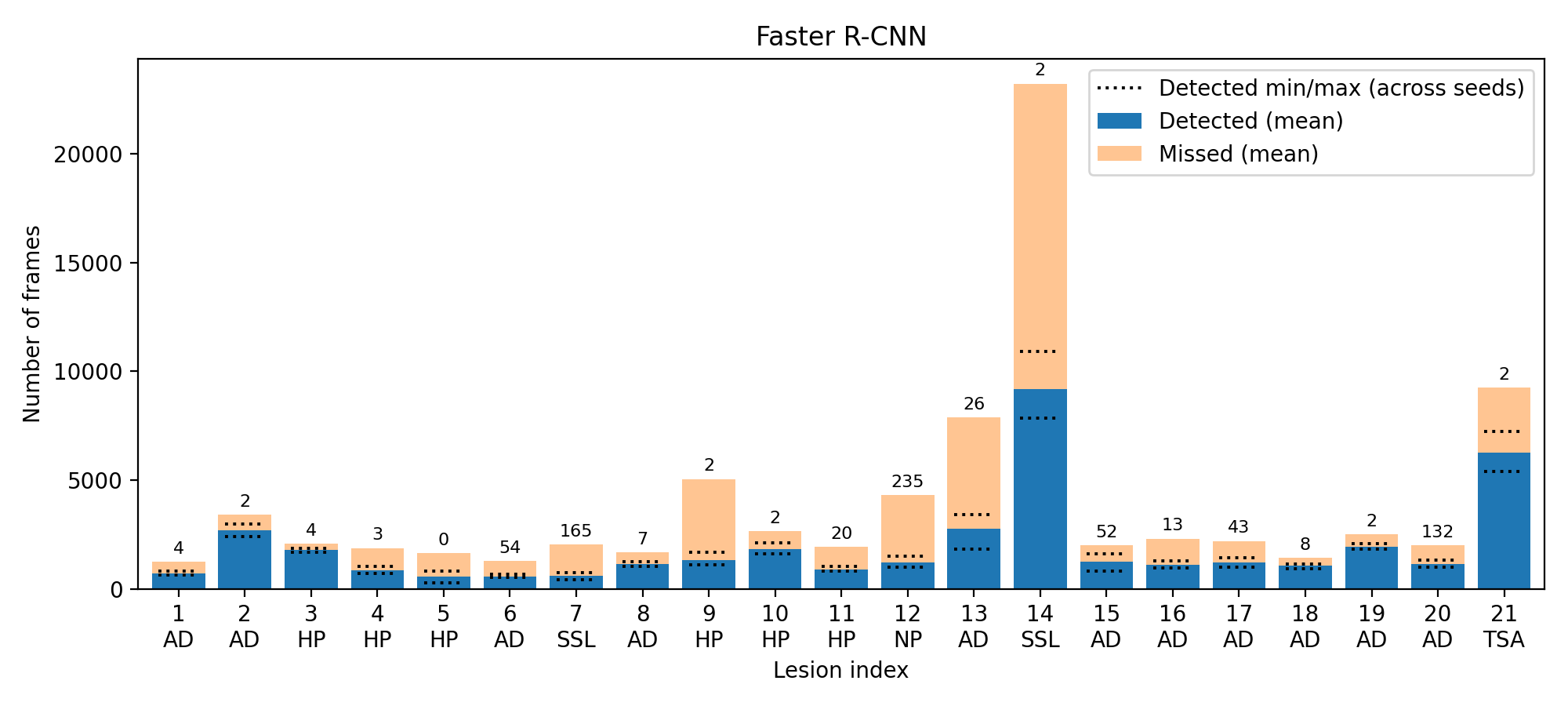}\\[2pt]
\includegraphics[width=0.87\linewidth]{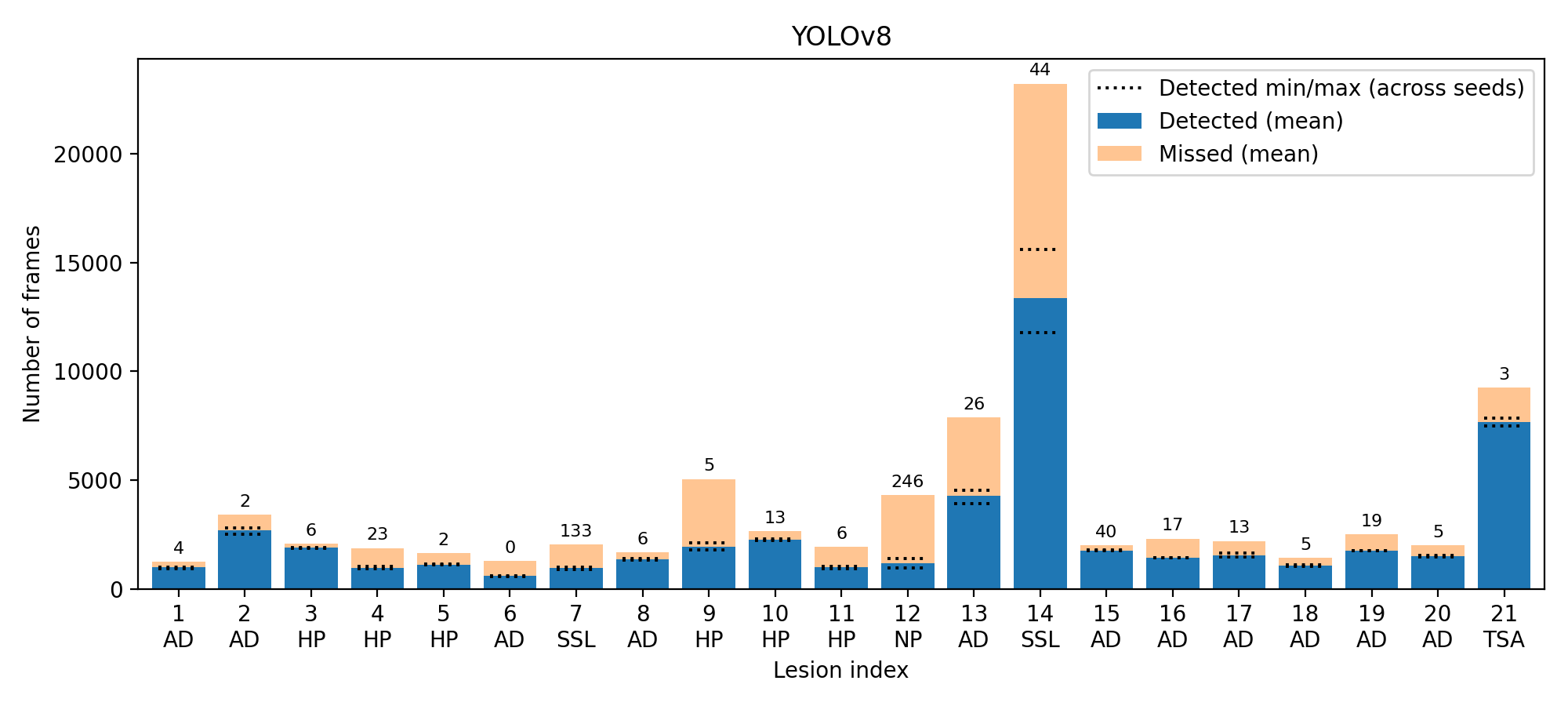}\\[2pt]
\includegraphics[width=0.87\linewidth]{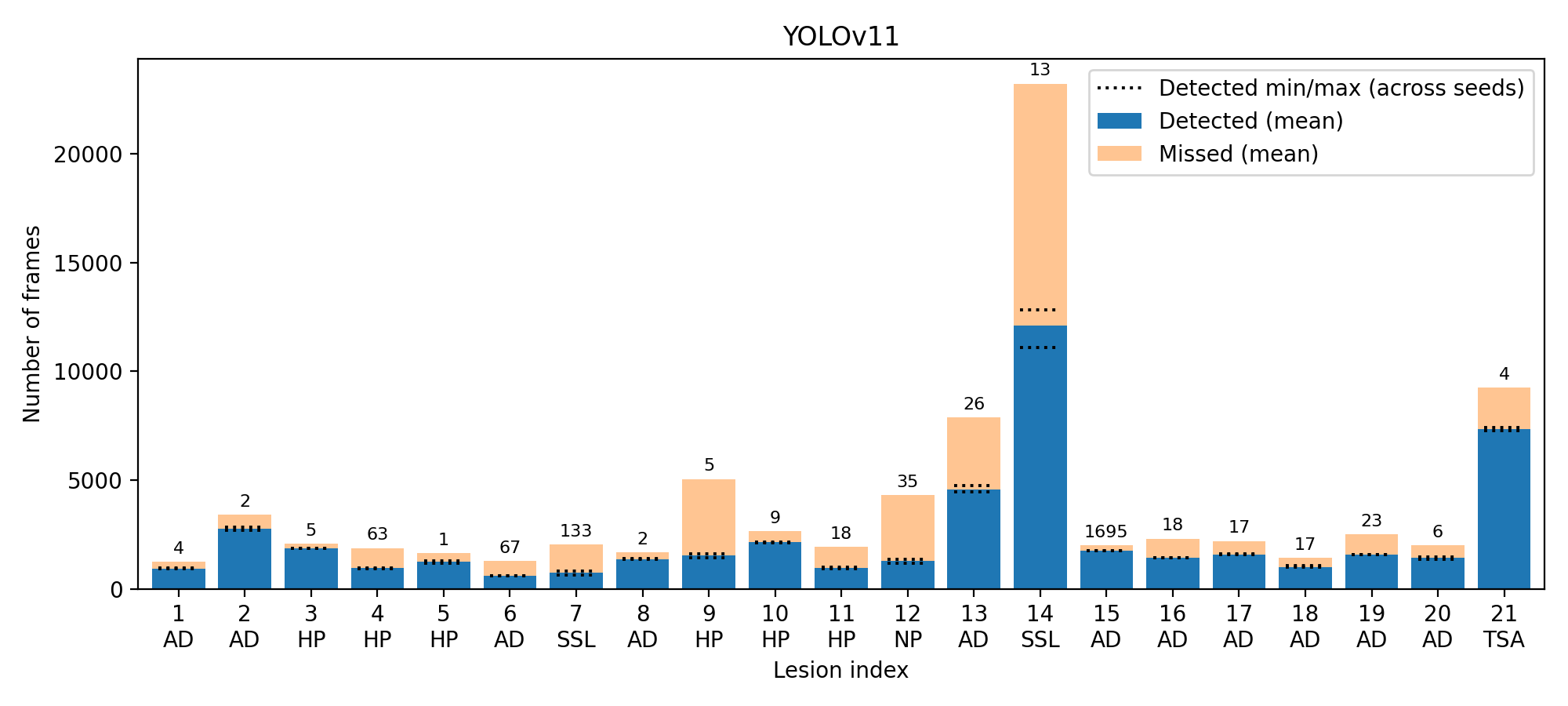}\\[2pt]
\includegraphics[width=0.87\linewidth]{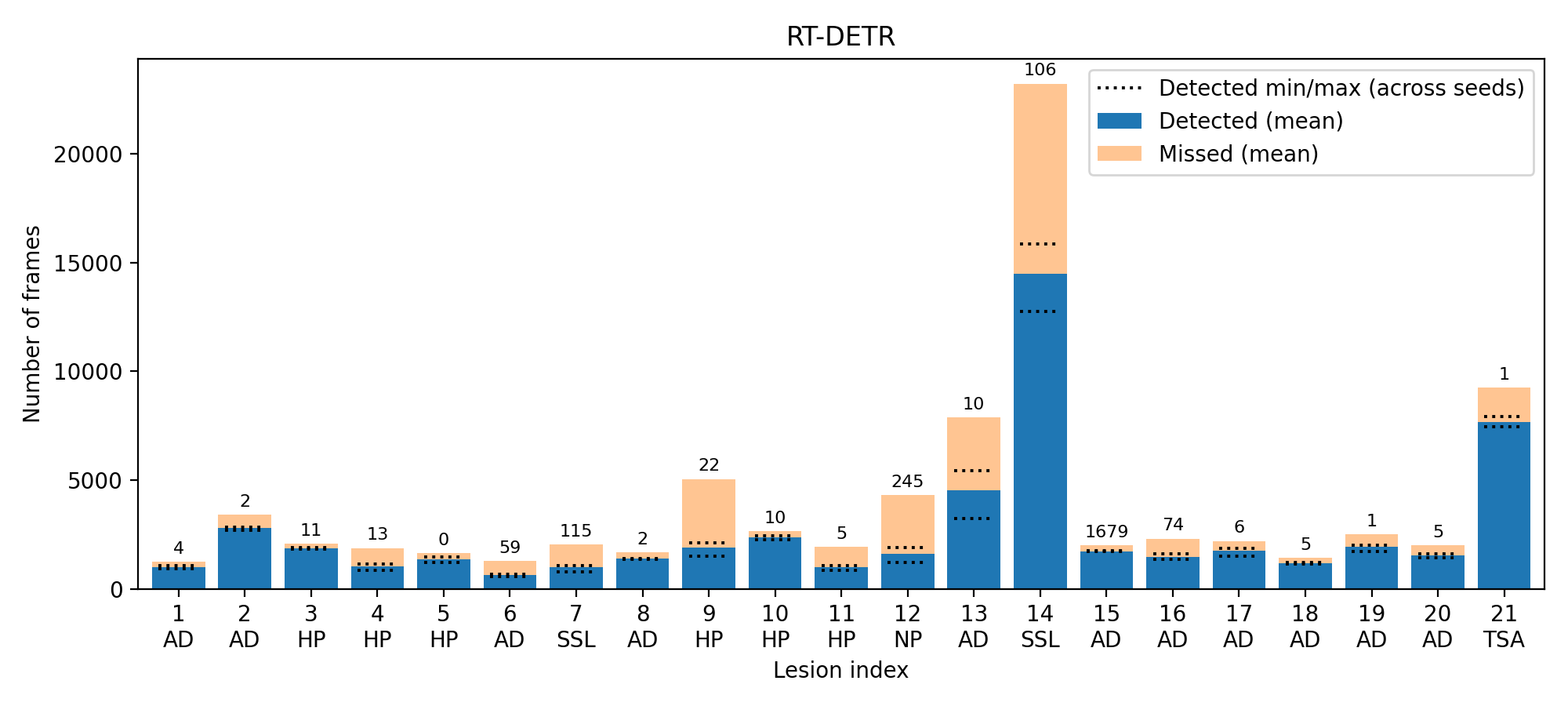}
\caption{Per-lesion detection persistence on REAL-Colon at
         model-specific operating thresholds.
         \textbf{Blue bars:} mean detected frames.
         \textbf{Orange bars:} mean missed frames.
         \textbf{Dotted lines:} min/max across seeds.
         The number above each bar is the first-detection frame index.
         Persistence varies widely by lesion, and the zero-shot SSL/TSA
         lesions are among the least consistently detected.}
\label{fig:sup-lesion-bars}
\end{figure}

\begin{figure}[htbp]
\centering
\includegraphics[width=0.52\linewidth]{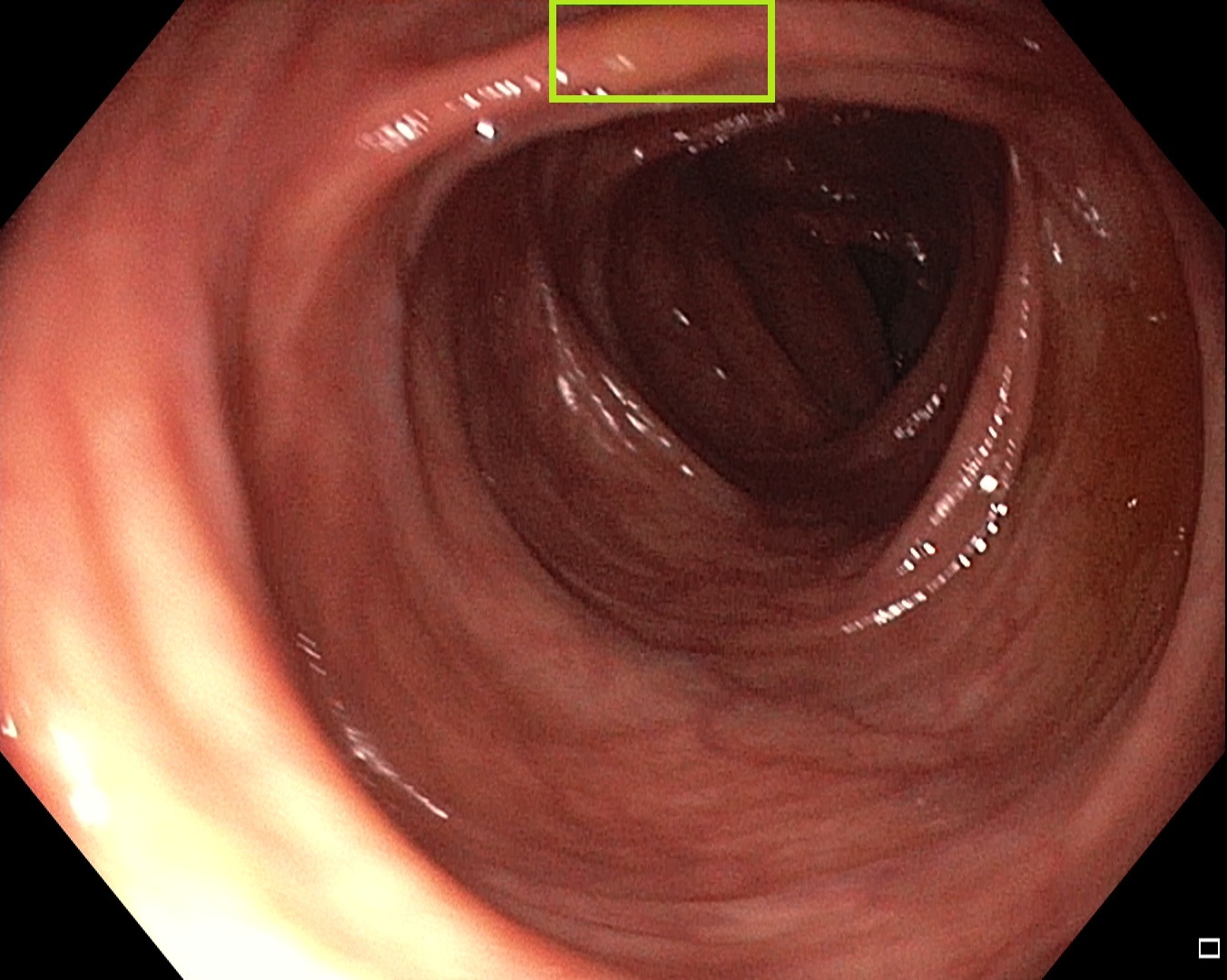}\\[4pt]
{\small\textbf{Lesion~7 (SSL)}}\\[8pt]
\includegraphics[width=0.52\linewidth]{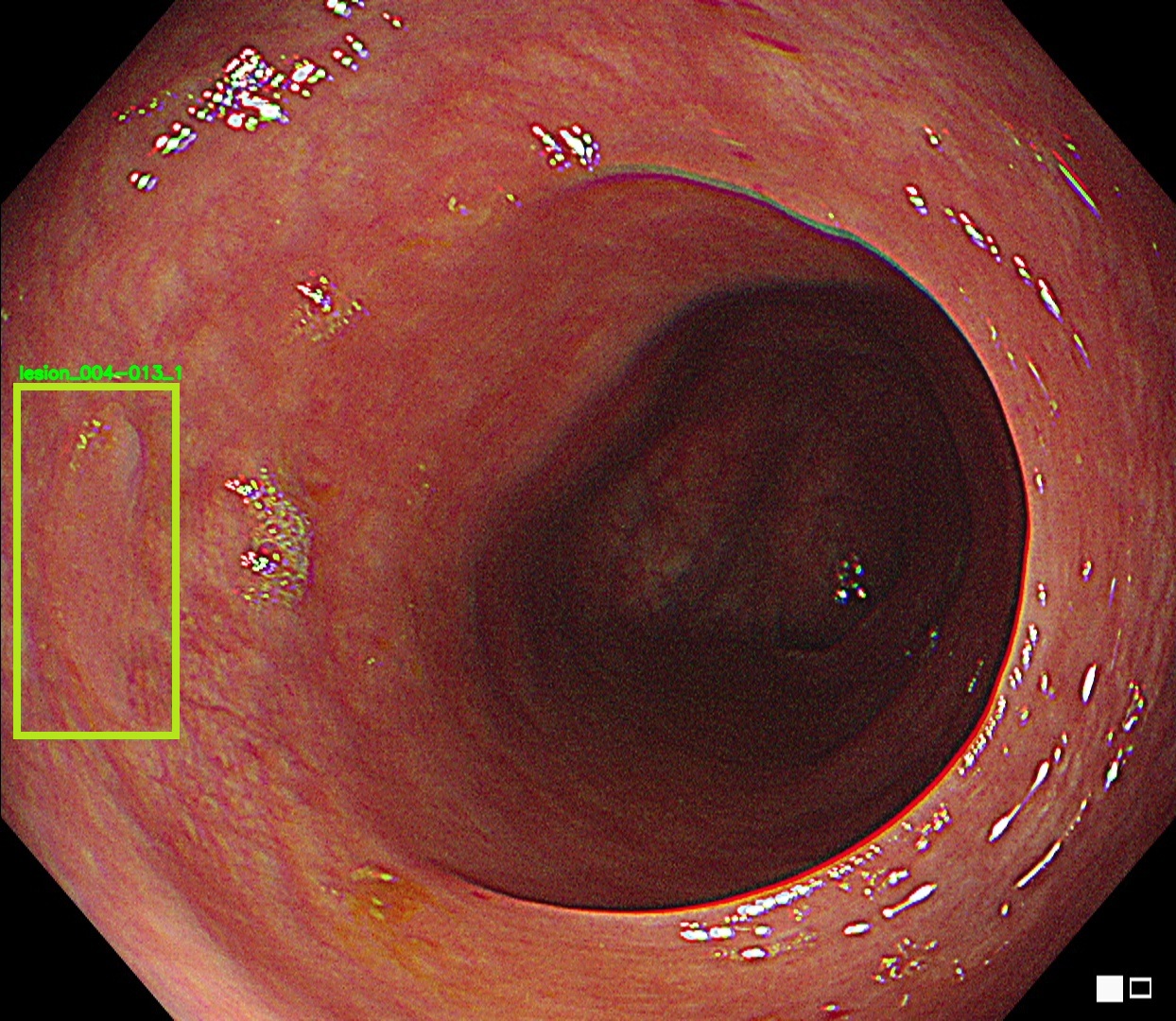}\\[4pt]
{\small\textbf{Lesion~14 (SSL)}}\\[8pt]
\includegraphics[width=0.52\linewidth]{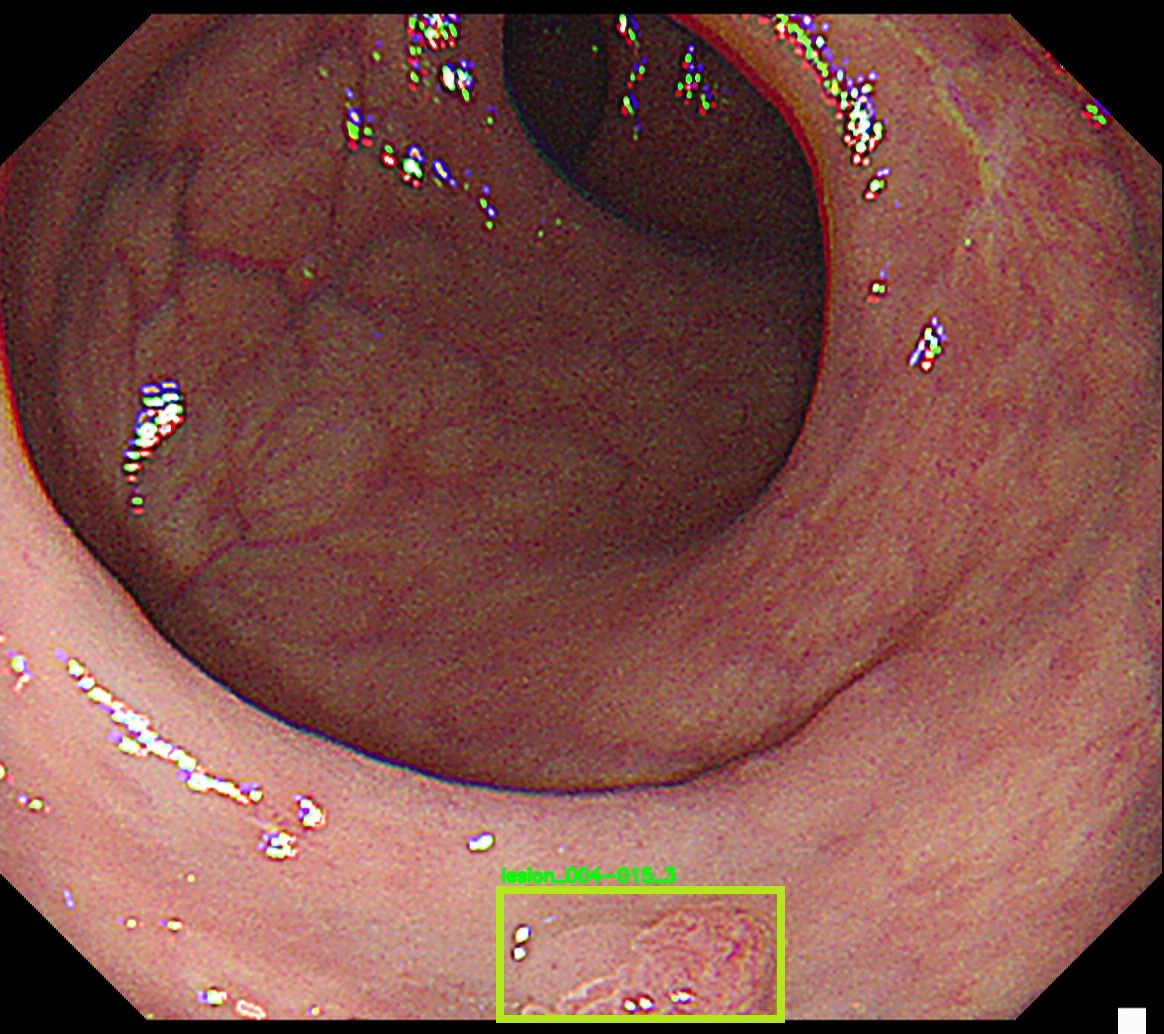}\\[4pt]
{\small\textbf{Lesion~21 (TSA)}}
\caption{Representative frames for the zero-shot histological
         subtypes in the test set.
         Top: Lesion~7 (SSL). Middle: Lesion~14 (SSL).
         Bottom: Lesion~21 (TSA).
         Green boxes indicate ground-truth annotations.
         These subtypes are absent from the training set; Lesion~14
         (SSL) shows the highest miss rates across all architectures.}
\label{fig:sup-ssl-examples}
\end{figure}

Detection-level and frame-level metrics do not capture whether a CADe
system finds each lesion, holds onto it, and flags it early enough to be
useful. To assess this temporal reliability, and to support the
lesion-level block of Table~\ref{tab:detection}, we track for each of the
$n=21$ REAL-Colon test lesions whether it is ever detected, the fraction
of its visible frames that are detected, and the latency to first
detection. A lesion counts as detected within a $1$, $3$, or $5$\,s window
if at least 15 of its frames in that window are detected; because
REAL-Colon's native frame rate varies across videos, these second-based
windows assume a nominal $30$\,fps and are approximate.
Two histological subtypes occur only in the test set and are therefore
zero-shot targets for every model: sessile serrated lesions (SSL, indices
7 and 14) and one traditional serrated adenoma (TSA, index 21).
Table~\ref{tab:sup-lesion} summarizes consistency and latency,
Figure~\ref{fig:sup-lesion-bars} shows the per-lesion detected-frame
fraction with first-detection indices, and
Figure~\ref{fig:sup-ssl-examples} shows representative frames for the
zero-shot lesions.

\subsubsection{Results}
All four detectors eventually match every lesion ($\ge1$ detected frame),
but they differ sharply in persistence and latency. At each detector's
operating threshold, RT-DETR detects $17.3/21$ lesions on more than half
their frames and fires earliest (first-frame latency $38.3$ frames),
whereas the convolutional detectors clear the $50\%$-persistence bar on at
most $16/21$ lesions and are slower to fire (Faster\,R-CNN $63.9$ frames).
The zero-shot subtypes are the hardest cases: Lesion~14 (SSL) shows the
highest miss rate across all architectures, consistent with its absence
from training. RT-DETR's earlier and more persistent detection is the
lesion-level counterpart of its operating-point advantage in
Table~\ref{tab:detection}.

\begin{table}[htbp]
\centering
\caption{Lesion-level consistency and latency on REAL-Colon
         ($n=21$ lesions, means over three seeds), under a fixed
         $\tau=0.20$ and under model-specific thresholds.
         Rows report how many lesions are detected at all and on
         ${\ge}25\%$/${\ge}50\%$ of their visible frames, how many are
         first detected within $1$/$3$/$5$\,s, and the latency to the
         first detected frame.
         RT-DETR detects lesions more persistently and earlier than the
         convolutional detectors.}
\label{tab:sup-lesion}
\setlength{\tabcolsep}{4pt}
\renewcommand{\arraystretch}{1.15}
\resizebox{\linewidth}{!}{%
\begin{tabular}{lcccc}
\toprule
 & \textbf{Faster R-CNN} & \textbf{YOLOv8} & \textbf{YOLOv11} & \textbf{RT-DETR} \\
\midrule
\multicolumn{5}{l}{\emph{Fixed confidence threshold $\tau=0.20$}} \\
\midrule
Lesions detected ($\ge1$ match)      & 21.0   & 21.0   & 21.0   & 21.0   \\
Lesions detected ($\ge25\%$ frames)  & 19.0   & 19.3 & 18.7 & 21.0   \\
Lesions detected ($\ge50\%$ frames)  & 11.7 & 12.7 & 12.7 & 19.7 \\
Detected within 1\,s                 & \phantom{1}1.0    & \phantom{1}1.7  & \phantom{1}1.0    & \phantom{1}7.7  \\
Detected within 3\,s                 & \phantom{1}7.0    & \phantom{1}5.3  & \phantom{1}5.3  & 12.0 \\
Detected within 5\,s                 & 11.0 & \phantom{1}8.0  & \phantom{1}7.7  & 15.3 \\
Latency first frame [frames]         & $63.9 {\scriptstyle \pm 42.5}$ & $82.1 {\scriptstyle \pm 30.6}$ & $103.0 {\scriptstyle \pm 37.6}$ & $12.5 {\scriptstyle \pm 7.2}$ \\
\midrule
\multicolumn{5}{l}{\emph{Model-specific thresholds (FPR\,$\approx$\,4--5\%)}} \\
\midrule
Lesions detected ($\ge1$ match)      & 21.0   & 21.0   & 21.0   & 21.0   \\
Lesions detected ($\ge25\%$ frames)  & 19.0   & 20.7 & 21.0   & 21.0   \\
Lesions detected ($\ge50\%$ frames)  & 11.7 & 16.0 & 15.7 & 17.3 \\
Detected within 1\,s                 & \phantom{1}1.0    & \phantom{1}3.7  & \phantom{1}3.0    & \phantom{1}4.7  \\
Detected within 3\,s                 & \phantom{1}7.0    & \phantom{1}8.7  & \phantom{1}7.3  & 10.0   \\
Detected within 5\,s                 & 11.0   & 12.3 & 10.7 & 12.7 \\
Latency first frame [frames]         & $63.9 {\scriptstyle \pm 42.5}$ & $46.8 {\scriptstyle \pm 26.3}$ & $57.3 {\scriptstyle \pm 26.3}$ & $38.3 {\scriptstyle \pm17.3}$ \\
\bottomrule
\end{tabular}%
}
\end{table}

\subsection{Runtime and Real-Time Suitability}
\label{sec:sup-runtime}
A CADe detector is only usable if it keeps pace with the video stream, so
alongside accuracy we measure end-to-end throughput. We time each
component (pre-processing, inference, post-processing) at batch size~1 on
a single NVIDIA\,H100\,NVL GPU, and report throughput from inference alone
($\mathrm{FPS}_\text{inf}$) and end to end ($\mathrm{FPS}_\text{total}$).
Real-time operation is conventionally defined as
${\ge}30$\,FPS~\cite{pascal_2}.

\subsubsection{Results}
All four detectors run in real time on an H100
(Table~\ref{tab:sup-runtime}). The YOLO models are fastest end to end
(YOLOv8 $181.8$\,FPS, YOLOv11 $163.9$\,FPS), Faster\,R-CNN sits just above
the real-time floor ($37.5$\,FPS), and RT-DETR lies between them
($51.0$\,FPS) while carrying the highest accuracy. RT-DETR therefore buys
its localization and early-detection advantage at roughly a $3\times$
throughput cost relative to YOLOv11, yet still comfortably clears the
$30$\,FPS bar.

\begin{table}[htbp]
\centering
\scriptsize
\caption{Runtime on REAL-Colon at batch size~1 on one NVIDIA\,H100\,NVL
         GPU (mean\,$\pm$\,std). Times are per image for pre-processing,
         inference, and post-processing. $\mathrm{FPS}_\text{inf}$ counts
         inference only and $\mathrm{FPS}_\text{total}$ the full pipeline.
         All detectors exceed the $30$\,FPS real-time threshold.}
\label{tab:sup-runtime}
\setlength{\tabcolsep}{4pt}
\renewcommand{\arraystretch}{1.10}
\resizebox{\linewidth}{!}{%
\begin{tabular}{lccccc}
\multicolumn{6}{l}{\textbf{Detectron2}} \\
\toprule
\textbf{Model} & \textbf{Inf [ms]} & \textbf{Eval [ms]} &
\textbf{Total [ms]} & \textbf{FPS$_\text{inf}$} &
\textbf{FPS$_\text{total}$} \\
\midrule
Faster R-CNN & $25.07 {\scriptstyle \pm 2.72}$ & $0.20 {\scriptstyle \pm 0.00}$ & $26.70 {\scriptstyle \pm 2.52}$ & 39.9 & 37.5 \\
\bottomrule
\end{tabular}%
}
\vspace{8pt}

\resizebox{\linewidth}{!}{%
\begin{tabular}{lcccccc}
\multicolumn{7}{l}{\textbf{Ultralytics}} \\
\toprule
\textbf{Model} & \textbf{Pre [ms]} & \textbf{Inf [ms]} &
\textbf{Post [ms]} & \textbf{Total [ms]} &
\textbf{FPS$_\text{inf}$} & \textbf{FPS$_\text{total}$} \\
\midrule
YOLOv8   & $0.57 {\scriptstyle \pm 0.06}$ & \phantom{1}$4.53 {\scriptstyle \pm 0.38}$  & $0.40 {\scriptstyle \pm 0.00}$ & \phantom{1}$5.50 {\scriptstyle \pm 0.44}$  & 220.7 & 181.8 \\
YOLOv11  & $0.50 {\scriptstyle \pm 0.00}$ & \phantom{1}$5.20 {\scriptstyle \pm 0.17}$  & $0.40 {\scriptstyle \pm 0.00}$ & \phantom{1}$6.10 {\scriptstyle \pm 0.17}$  & 192.3 & 163.9 \\
RT-DETR  & $0.50 {\scriptstyle \pm 0.00}$ & $18.87 {\scriptstyle \pm 0.29}$ & $0.23 {\scriptstyle \pm 0.06}$ & $19.60 {\scriptstyle \pm 0.35}$ & \phantom{1}53.0  & \phantom{1}51.0  \\
\bottomrule
\end{tabular}%
}
\end{table}

\subsection{Ablation Studies}
\label{sec:sup-ablation}
To ground the training and inference choices used in the main paper, we
ablate the main design and training parameters on YOLOv11: input
resolution, model capacity, negative-frame sampling, and optimizer.
Every run uses model-specific confidence thresholds chosen for comparable
frame-level FPR under the $\mathrm{IoU}>0$ criterion.

\subsubsection{Input Resolution}
\label{sec:sup-abl-resolution}
Input resolution trades detection accuracy against throughput and sets the
point of comparison with the original REAL-Colon baseline. We evaluate
YOLOv11-M at $224\times224$, $640\times640$, and $1024\times1024$
(Table~\ref{tab:sup-abl-res}), and add a $300\times300$ run
(Table~\ref{tab:sup-abl-300}) matching the original author
setup~\cite{realColon}; because that setup does not specify the IoU and
confidence thresholds behind its frame-level metrics, its TPR and FPR are
shown for observational reference only.

\paragraph{Results.}
The $640\times640$ setting gives the best accuracy-to-throughput trade-off
and is the resolution used in the main paper. Scaling to $1024\times1024$
adds only $+0.016$ $\mathrm{mAP}_{50}$ while dropping end-to-end throughput
from $163.9$ to $143.8$\,FPS, and $224\times224$ loses accuracy for little
further speed. At the baseline-matched $300\times300$, all four detectors
land close to the original SSD-300 reference ($\mathrm{mAP}_{50}=0.338$),
with RT-DETR highest ($0.432$).
\begin{table}[htbp]
\centering
\caption{Effect of input resolution on YOLOv11-M (REAL-Colon).
         $640\times640$ gives the best accuracy-to-throughput trade-off and
         is the resolution used in the main paper.}
\label{tab:sup-abl-res}
\small
\setlength{\tabcolsep}{5pt}
\renewcommand{\arraystretch}{1.15}
\begin{tabular}{lccccccc}
\toprule
Resolution & conf &
$\mathrm{mAP}_{50:95}$ & $\mathrm{mAP}_{50}$ &
TPR & FPR &
$\mathrm{FPS}_\text{inf}$ & $\mathrm{FPS}_\text{total}$ \\
\midrule
$224\times224$   & 0.01 & 0.184 & 0.289 & 0.444 & 0.044 & 212.9 & 183.4 \\
$640\times640$   & 0.05 & 0.246 & 0.384 & 0.584 & 0.043 & 192.3 & 163.9 \\
$1024\times1024$ & 0.03 & 0.255 & 0.400 & 0.576 & 0.044 & 184.3 & 143.8 \\
\bottomrule
\end{tabular}
\end{table}

\begin{table}[htbp]
\centering
\caption{Comparison at $300\times300$ following the original REAL-Colon
         author setup~\cite{realColon} on the full dataset. Because the
         original IoU and confidence thresholds are unspecified, TPR and
         FPR are for observational reference only.}
\label{tab:sup-abl-300}
\small
\setlength{\tabcolsep}{6pt}
\renewcommand{\arraystretch}{1.15}
\begin{tabular}{lccccc}
\toprule
Model & conf & $\mathrm{mAP}_{50:95}$ & $\mathrm{mAP}_{50}$ & TPR & FPR \\
\midrule
Faster R-CNN            & 0.200 & 0.203 & 0.330 & 0.497 & 0.043 \\
YOLOv8-M                & 0.013 & 0.221 & 0.342 & 0.524 & 0.048 \\
YOLOv11-M               & 0.015 & 0.194 & 0.297 & 0.504 & 0.046 \\
RT-DETR                 & 0.250 & 0.290 & 0.432 & 0.634 & 0.050 \\
SSD-300 (author setup)  &  --  & 0.216 & 0.338 & 0.505 & 0.054 \\
\bottomrule
\end{tabular}
\end{table}

\subsubsection{Model Capacity, Negative Sampling, and Optimizer}
\label{sec:sup-abl-hyper}
We next isolate three training choices on YOLOv11: model size (S/M/L), the
ratio of negative to positive frames (1:0.15, 1:0.50, 1:1.00), and the
optimizer (Adam, AdamW, SGD), holding the rest of the pipeline fixed
(Table~\ref{tab:sup-abl-hyper}). Negative sampling is the choice most
specific to full-procedure data, where negative frames vastly outnumber
positives.

\paragraph{Results.}
Accuracy grows with capacity ($\mathrm{mAP}_{50}$: S $0.353$, M $0.384$,
L $0.420$). The main paper reports the M variant as the
accuracy-to-throughput compromise matching the $640\times640$ operating
point. A balanced $1{:}1$ negative ratio is best
($\mathrm{mAP}_{50}=0.401$): starving the model of background forces a much
higher confidence threshold ($\tau=0.27$) to hold false positives down,
which costs sensitivity. Optimizer choice barely moves the result at
$224\times224$ ($\mathrm{mAP}_{50}$ within $0.285$--$0.289$), so we keep
the default. These runs fix the $1{:}1$ negative sampling and the M-size
model used throughout the main experiments.

\begin{table}[htbp]
\centering
\caption{Effect of model capacity, negative-sampling ratio, and optimizer
         on YOLOv11 (REAL-Colon, $640\times640$ unless noted).
         A balanced $1{:}1$ negative ratio and larger capacity help most,
         while the optimizer barely matters.}
\label{tab:sup-abl-hyper}
\small
\setlength{\tabcolsep}{6pt}
\renewcommand{\arraystretch}{1.15}
\begin{tabular}{lccccc}
\toprule
Model / Config & conf &
$\mathrm{mAP}_{50:95}$ & $\mathrm{mAP}_{50}$ & TPR & FPR \\
\midrule
YOLOv11-S        & 0.040 & 0.221 & 0.353 & 0.570 & 0.045 \\
YOLOv11-M        & 0.050 & 0.246 & 0.384 & 0.584 & 0.043 \\
YOLOv11-L        & 0.007 & 0.276 & 0.420 & 0.628 & 0.044 \\
\midrule
Neg.\ ratio 1:0.15 & 0.27 & 0.209 & 0.328 & 0.482 & 0.048 \\
Neg.\ ratio 1:0.50 & 0.27 & 0.155 & 0.254 & 0.450 & 0.047 \\
Neg.\ ratio 1:1.00 & 0.10 & 0.253 & 0.401 & 0.602 & 0.048 \\
\midrule
Adam             & 0.005 & 0.181 & 0.285 & 0.474 & 0.050 \\
AdamW            & 0.070 & 0.182 & 0.285 & 0.500 & 0.048 \\
SGD              & 0.050 & 0.184 & 0.289 & 0.444 & 0.044 \\
\bottomrule
\end{tabular}
\end{table}

\subsection{Comparison with Commercial CADe Systems}
\label{sec:sup-clinical}
To place the open detectors' operating points in the context of deployed
technology, we compare against frame-level metrics for commercial CADe
systems and the open-source EndoMind detector, as reported by Troya et
al.~\cite{endomind} (Table~\ref{tab:sup-commercial}). The reference set
spans two generations of the GI Genius system (v1, v2), two EndoAID
configurations (A, B), and EndoMind, an open-source real-time detector.
These figures come from a different dataset, annotation protocol, and
evaluation pipeline, so the comparison is contextual rather than a
head-to-head benchmark.
\begin{table}[htbp]
\centering
\caption{Frame-level performance of commercial CADe systems and the
         open-source EndoMind detector ($\mathrm{IoU}>0$, $\tau=0.2$), as
         reported in~\cite{endomind}. Values come from a different dataset
         and pipeline, so they contextualize rather than directly compare
         with our detectors.}
\label{tab:sup-commercial}
\setlength{\tabcolsep}{5pt}
\renewcommand{\arraystretch}{1.15}
\resizebox{\linewidth}{!}{%
\begin{tabular}{lccccc}
\toprule
\textbf{Metric} &
\textbf{GI Genius v1} & \textbf{GI Genius v2} &
\textbf{EndoAID A}    & \textbf{EndoAID B}    &
\textbf{EndoMind} \\
\midrule
Sensitivity / TPR        & 0.5063 & 0.6785 & 0.6560 & 0.5295 & 0.6022 \\
Specificity              & 0.9694 & 0.9577 & 0.9719 & 0.9930 & 0.9589 \\
FPR                      & 0.0275 & 0.0380 & 0.0252 & 0.0063 & 0.0369 \\
Precision                & 0.5966 & 0.5701 & 0.6658 & 0.8720 & 0.5499 \\
$F_1$ score              & 0.5915 & 0.6391 & 0.6943 & 0.7177 & 0.5960 \\
First detection (ms)     & 1510   & 607    & 659    & 1316   & 1083   \\
\bottomrule
\end{tabular}%
}
\end{table}

\subsubsection{Results}
The open detectors evaluated here reach sensitivities of $0.60$--$0.72$,
matching or exceeding the commercial range ($0.51$--$0.68$), but at
somewhat higher false-positive rates (${\approx}0.045$ vs.\
${\approx}0.030$). Their first-detection latency ($1.2$--$2.1$\,s at
model-specific thresholds, at the same nominal $30$\,fps) is broadly in
line with the commercial range ($0.6$--$1.5$\,s). Given the differing
hardware and pipelines, the takeaway is qualitative: open detectors
already operate in the same sensitivity and latency envelope as commercial
CADe, with headroom mainly in false-positive control.

\subsection{Qualitative Detection Examples}
\label{sec:sup-qualitative}
\begin{figure}[htpb]
\centering
\includegraphics[width=\linewidth]{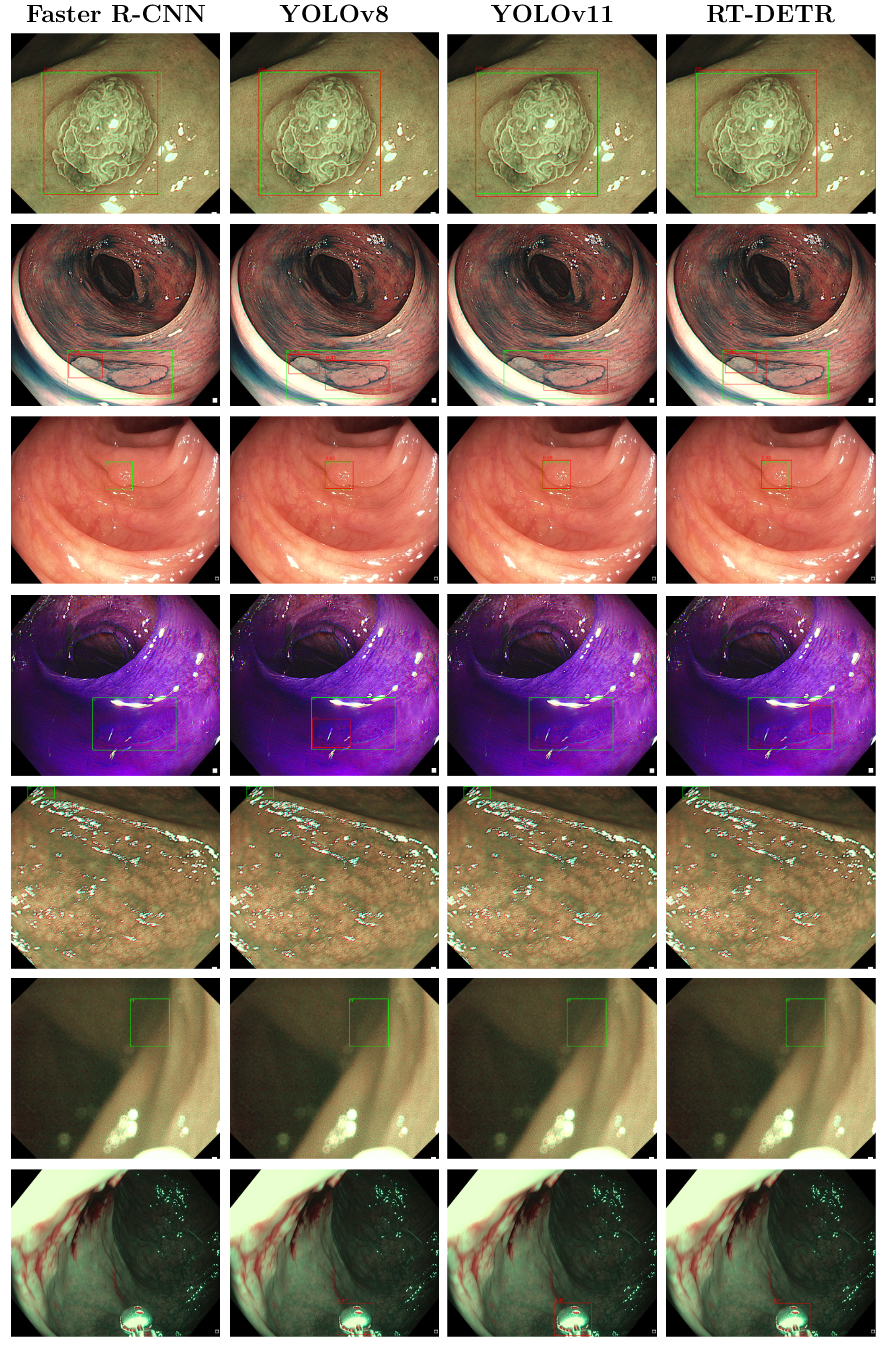}
\caption{Qualitative detection results across all four architectures on
         representative REAL-Colon test frames. Each \textbf{row} is one
         frame under varying visual conditions. The \textbf{columns} are
         Faster\,R-CNN, YOLOv8, YOLOv11, and RT-DETR (left to right).
         Green boxes are ground-truth annotations and red boxes are model
         predictions. Predictions agree on clear polyps and diverge on
         flat and NBI frames.}
\label{fig:sup-summary-grid}
\end{figure}

Aggregate metrics hide where the detectors agree and where they fail
together. To make the failure modes concrete, we show side-by-side
predictions from all four architectures on representative REAL-Colon test
frames, spanning clearly visible polyps, NBI-stained mucosa, flat lesions,
and specular highlights (Figure~\ref{fig:sup-summary-grid}), together with
frames that defeat every model (Figure~\ref{fig:sup-challenges}).

\subsubsection{Results}
\begin{figure}[htbp]
\centering
\includegraphics[width=\linewidth]{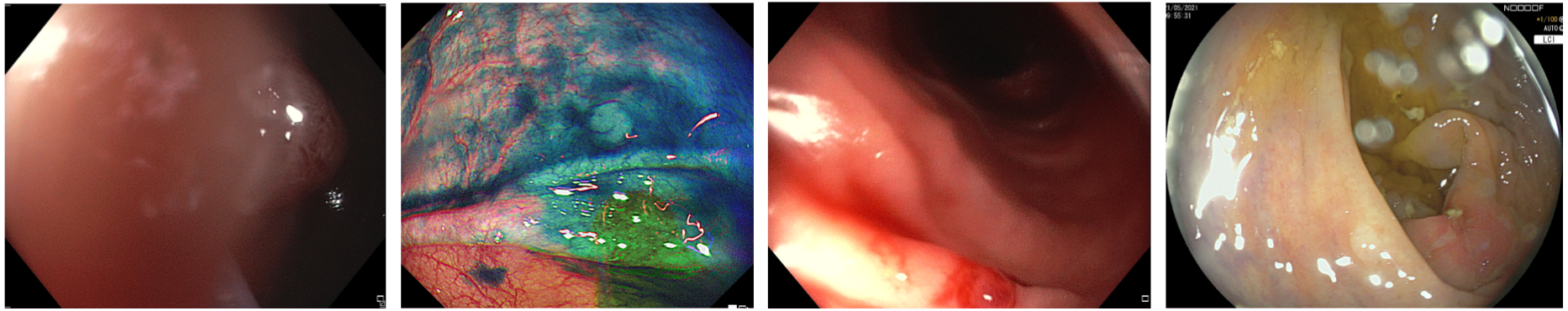}
\caption{Challenging REAL-Colon test frames on which all evaluated
         architectures produce false negatives or false positives.
         From left to right: specular highlights, motion blur,
         instrument occlusion, and a flat non-polypoid lesion.
         These conditions are the main shared failure mode across
         detectors.}
\label{fig:sup-challenges}
\end{figure}

On clearly visible polyps all four detectors localize the lesion, and
their predictions diverge mainly on flat, non-polypoid, and NBI frames.
The shared failure cases (Figure~\ref{fig:sup-challenges}) are dominated
by specular highlights, motion blur, instrument occlusion, and flat
lesions, the same conditions behind the near-zero small-polyp AP and the
low-persistence lesions above. No architecture is robust to these frames,
which marks the main headroom for full-procedure detection.

\newpage
\section{Evaluation Metric Definitions}
\label{sec:sup-metrics}
\setcounter{table}{0}\setcounter{figure}{0}\setcounter{equation}{0}
This section formally defines every metric reported in the main paper and
above. Object detection performance can be evaluated under different
formulations depending on how detector outputs are interpreted.
\emph{Detection-level} metrics assess localization accuracy of individual
bounding boxes. \emph{Frame-level} metrics reformulate detection as binary
classification per frame, quantifying the false-alert burden on negative
video segments. Standard benchmarks such as COCO rely primarily on
detection-level metrics~\cite{coco}, whereas medical video analysis
additionally requires metrics that characterize temporal aggregation and
real-time feasibility.

\subsection{Detection-Level Metrics}
\label{sec:sup-metrics-det}
Individual predicted bounding boxes are matched to ground-truth (GT)
boxes using an Intersection-over-Union (IoU) criterion under a
one-to-one matching policy.
A prediction is a true positive (TP) if it matches a GT box above the IoU threshold. Unmatched predictions are false positives (FP),
and unmatched GT boxes are false negatives (FN).
True negatives are undefined at the detection level as the background does not form a finite set of negative instances~\cite{detection_metrics}.
Precision, recall, and mean Average Precision (mAP) follow standard
COCO definitions~\cite{coco,detection_metrics}:
\begin{equation}
  \text{Recall} = \frac{\mathrm{TP}}{\mathrm{TP}+\mathrm{FN}},\quad
  \text{Precision} = \frac{\mathrm{TP}}{\mathrm{TP}+\mathrm{FP}},\quad
  \mathrm{mAP} = \frac{1}{n}\sum_{k=1}^{n}\mathrm{AP}_{k}.
  \label{eq:det-metrics}
\end{equation}
Sensitivity is equivalent to recall.
The $F_\beta$ score weights precision and recall via the parameter
$\beta$. Setting $\beta=2$ prioritizes recall, which is appropriate for polyp detection where missed lesions carry greater clinical risk than false alarms~\cite{classification_metrics}:
\begin{equation}
  F_\beta = (1+\beta^2)
  \frac{\text{Precision}\cdot\text{Recall}}
       {\beta^2\,\text{Precision}+\text{Recall}}.
  \label{eq:fbeta}
\end{equation}

\subsection{FROC and AFROC}
\label{sec:sup-metrics-afroc}
The standard ROC assigns a single decision per image and therefore
does not distinguish between one false alarm and multiple false alarms
within the same image~\cite{froc_soup,froc_afroc2}.
The Free-response ROC (FROC)~\cite{froc_old} extends ROC by allowing
multiple detections per image, characterizing the trade-off between
sensitivity and the average number of false detections per image (FPPI):
\begin{equation}
  \mathrm{FPPI} =
  \frac{\text{number of false localizations}}
       {\text{number of images}}.
  \label{eq:fppi}
\end{equation}
Because FPPI is unbounded, the Alternative FROC (AFROC)~\cite{froc_old,froc_soup} replaces it with the
false-positive fraction (FPF), defined as the fraction of negative
images raising at least one false-positive alert above the operating
threshold:
\begin{equation}
  \mathrm{FPF} =
  \frac{\text{negative images with}\;\ge1\;\text{FP}}
       {\text{total negative images}}
  \;\in [0,1].
  \label{eq:fpf}
\end{equation}
This confines the curve to the unit square and supports ROC-like
scalar summaries such as the area under the AFROC curve, while
retaining sensitivity on the ordinate~\cite{froc_old,froc_afroc2}.

\subsection{Frame-Level Metrics}
\label{sec:sup-metrics-frame}
Detector outputs are collapsed to a binary alert per frame at
confidence threshold $\tau$~\cite{classification_metrics}.
A frame is predicted positive if it contains at least one detection
exceeding $\tau$ (and overlapping a ground-truth box at $\mathrm{IoU}>0$ when GT is present),
and negative otherwise.
This creates a standard confusion matrix where TN becomes defined:
\begin{itemize}
  \item $\mathrm{TP}_\text{frame}$: GT-positive frame with
        at least one valid detection,
  \item $\mathrm{FN}_\text{frame}$: GT-positive frame without
        a valid detection,
  \item $\mathrm{FP}_\text{frame}$: GT-negative frame with
        at least one detection,
  \item $\mathrm{TN}_\text{frame}$: GT-negative frame without
        any detection.
\end{itemize}
Frame-level TPR, FPR, and specificity are then:
\begin{equation}
  \mathrm{TPR} = \frac{\mathrm{TP}}{\mathrm{TP}+\mathrm{FN}},\quad
  \mathrm{FPR} = \frac{\mathrm{FP}}{\mathrm{FP}+\mathrm{TN}},\quad
  \text{Specificity} = \frac{\mathrm{TN}}{\mathrm{TN}+\mathrm{FP}}.
  \label{eq:frame-metrics}
\end{equation}

\subsection{Runtime Metrics}
\label{sec:sup-metrics-runtime}
End-to-end image latency and throughput are:
\begin{equation}
  t_\text{img} = t_\text{pre} + t_\text{inf} + t_\text{post},
  \qquad
  \mathrm{FPS} = \frac{1000}{t_\text{img}[\mathrm{ms}]}.
  \label{eq:fps}
\end{equation}
Real-time operation is conventionally defined as
$\ge30$\,FPS, corresponding to an end-to-end latency below
33\,ms~\cite{pascal_2}.

\end{document}